\documentclass[a4paper,aps,nofootinbib,onecolumn,notitlepage]{revtex4-1}
\RequirePackage[english]{babel}
\RequirePackage[latin1]{inputenc}
\RequirePackage[T1]{fontenc}
\RequirePackage{mathrsfs}
\RequirePackage{amsmath}
\RequirePackage{amssymb}
\RequirePackage{amsbsy}
\RequirePackage{bm}
\usepackage[lofdepth,lotdepth]{subfig}
\usepackage{graphicx}
\usepackage{multirow}
\usepackage{color} 
\usepackage{makecell} 
\usepackage{dsfont}
\def\beq{\begin{equation}}
\def\eeq{\end{equation}}

\def\g{\mathds G}
\def\fo{\mathcal F}

\def\ddemunu{_{;\mu\nu}}

\def\gd{g_{\mu\nu}}

\def\pa{\partial}

\def \A {{\mathbb A}}
\def \K {{\mathbb K}}

\def \Q {{\mathbb Q}}
\def \S {{\mathbb S}}
\def\case#1/#2{\frac{#1}{#2}}
\def\DerN#1{\frac{d #1}{d N}}

\def\rf#1{(\ref{#1})}

\begin{document}

\title{Cosmology of $f(R, \Box R)$ gravity}
\author{Sante Carloni}
\email{sante.carloni@gmail.com}
\affiliation{Centro de Astrofis\'ica e Gravita\c{c}\~ao - CENTRA,
Instituto Superior T\'{e}cnico - IST,
Universidade de Lisboa - UL,
Avenida Rovisco Pais 1, 1049-001, Portugal}

\author{Jo\~{a}o Lu\'{i}s Rosa}
\email{joaoluis92@gmail.com}
\affiliation{Centro de Astrofis\'ica e Gravita\c{c}\~ao - CENTRA,
Instituto Superior T\'{e}cnico - IST,
Universidade de Lisboa - UL,
Avenida Rovisco Pais 1, 1049-001, Portugal}

\author{Jos\'{e} P. S. Lemos}
\email{joselemos@ist.utl.pt}
\affiliation{Centro de Astrofis\'ica e Gravita\c{c}\~ao - CENTRA,
Instituto Superior T\'{e}cnico - IST,
Universidade de Lisboa - UL,
Avenida Rovisco Pais 1, 1049-001, Portugal}

\date{\today}

\begin{abstract}
Using dynamical system analysis, we explore the cosmology of theories
of order up to eight order of the form $f(R, \Box R)$. The phase space of these
cosmology reveals that higher-order terms can have a dramatic
influence on the evolution of the cosmology, avoiding the onset of
finite time singularities. We also confirm and extend some of results
which were obtained in the past for this class of theories.
\end{abstract}
\maketitle

\section{Introduction}
General relativity deals with second-order differential equations
for the metric $g_{\mu\nu}$.
Higher-order modifications of the gravitational interaction have been for long time the focus of intense investigation. They have been proposed for a number of reasons including the first attempts of unification of gravitation and  other fundamental interactions. Nowadays, the main reason why one considers this kind of extension in general relativity is of quantum origin. Studies on the renormalisation of the stress-energy tensor of quantum fields in the framework of a semi classical approach to genral relativity, i.e., what we call quantum field theory in curved spacetime, shows that such corrections are needed to take into account the differences between the gravitation of quantum fields and the gravitation of
classical fluids \cite{BOSbook,Birrell:1982ix}. 

With the introduction of the paradigm of inflation and the requirement of a field able to drive it, it was natural, although not obvious, to consider these quantum corrections as the engine of the inflationary mechanism. Starobinski \cite{Starobinsky:1980te} was able to show explicitly in the case of fourth-order corrections to general relativity that this was indeed the case: quantum corrections could induce an inflationary phase.  Such result should not be surprising. Fourth-order gravity carries an additional scalar  degree of freedom and this scalar degree of freedom can drive an inflationary phase. In the following years other researchers \cite{Amendola:1993bg,Gottlober:1989ww,Wands,Berkin-Maeda} tried to look at the behaviour of sixth-order corrections, to see if in this case one could obtain a richer inflationary phase and more specifically a cosmology with multiple inflationary phases. However, it turned out that this is not the case: in spite of the presence of an additional scalar degree of freedom, multiple inflationary phases were not possible. The reason behind this result is still largely unknown. 

The discovery of the dark energy offered yet another application for
the additional degree of freedom of higher-order gravity. Like in the
case of inflation, this perspective offered an elegant way to explain
dark energy: higher-order corrections were a geometrical way to
interpret the mysterious new component of the Universe
\cite{GeomDE}. Here an important point should be stressed: differently
from the standard perturbative investigation of a physical system, in
the case of higher-order gravity, the behaviour of the new theory
cannot be deduced as a small perturbation of the original second-order
one. The reason is that, since the equations of motion switch order,
the dynamics of the perturbed system are completely different from the
non-perturbed one whatever the (non-zero) value of the smallness
parameter.  For this reason, the properties of higher-order gravity
cannot be deduced from their lower-order counterpart, even if the
higher-order terms are suppressed by a small coupling constant. This
fact calls for a complete reanalysis of the phenomenology of these
theories. Such study should be performed with tools designed
specifically for this task, which therefore contain no hidden
assumptions or priors which might compromise the final result. One of
these tools, which will be used in the following is the so called
Dynamical System Approach (DSA)
\cite{Collins,DSA Book}. DSA has been used now for long time to
understand the dynamics of cosmologies of a number of different
modifications of general relativity (see
e.g. Refs.~\cite{Bahamonde:2017ize,Odintsov:2017tbc,Carloni:2017ucm,Carloni:2015jla,Carloni:2014pha,Carloni:2009jc,Alho:2016gzi,Carloni:2015lsa,Carloni:2015bua,Carloni:2013hna,Bonanno:2011yx,Roy:2011za,Carloni:2009nc,Carloni:2008jy,Abdelwahab:2007jp,Carloni:2007br,Carloni:2007eu,Carloni:2006mr,Leach:2006br,Carloni:2004kp}). It
is based on the definition of a set of expansion normalised variables
of clear physical meaning which help the physical interpretation of
the orbits obtained. The first attempt to apply this technique to  a theory of order six was made in Ref.~\cite{Berkin-Maeda}. Recently a new version of DSA has been
proposed \cite{Carloni:2015jla}, which helped clarify the cosmological dynamics of
fourth-order gravity, revealing new aspects of these theories. The
technique is also extendable to consider higher-order theories and in
the following we will propose a formalism able to treat a subclass of
theories of gravity of order six and eight. 

Among the many unresolved issues that are known to affect higher-order theories, it is worth to mention briefly the so-called Ostrogradski theorem \cite{Woodard:2006nt}. The theorem shows that for a generic system with a higher-order Lagrangian, there exist a conserved quantity $H$ corresponding to time shift invariance. When this quantity is interpreted as a Hamiltonian, by the definition of a suitable Legendre transformation, it can be shown that such Hamiltonian, not being limited from below, leads to the presence of undesirable features of the theory upon quantisation, whereas the classical behaviour, which includes classical cosmology, has no problem. In view of this conclusion higher-order theories, with the notable exception of $f(R)$ gravity, are deemed as unphysical. The most important issue for this work is then, why bother with higher-order gravity? We can give two arguments. The first is that, as mentioned above, the higher-order terms we will consider are terms of a series of corrections arising in a renormalisation procedure. In this perspective, therefore, there is no requirement that the truncated series had the same convergence property of its sum. A typical example is the Taylor series of $\sin(x)$. The truncated series is not bound, whereas its full sum is. In the same way the truncation of the original semiclassical model that gives rise to a higher-order theory might be fundamentally flawed on the quantum point of view. The problem only arises if one chooses the {\em complete} theory of quantum gravity to be given by a $n$-order truncation.  The second is that a study of the behaviour of the truncation allows an understanding of the interplay between the different terms of the development and in particular if and how the pathologies of the theory at a certain order are changed by the terms of higher-order. This  on one hand allows to give statements on the validity of the procedure of renormalisation in quantum field theory in curved spacetime and, on the other hand, it is  interesting in the context of the cosmology of fourth-order gravity, as it is known that this class of theories can present a number of issues, i.e., scale factor can evolve towards a singularity at finite time
\cite{Carloni:2015jla}
which is independent from the Ostrogradski instability. An analysis of the higher-order theories can therefore shed light on the real nature of these pathologies.

In this paper we will propose a DSA able to give a description of the dynamics of cosmological models based on a subclass of theories of gravity represented by the Lagrangian density $\mathcal L=f(R, \Box R)$. We will show that the higher-order terms in these theories act in an unexpected way on the cosmology: They can be dominant and prevent the appearance of finite time singularities. The calculations involved in this task are formidable so we will give the full expression only when strictly necessary.

The paper is organised as follows. In Section II, we give the general form of the field equations. In Section III, we construct a general formalism for theories of order six and we consider two specific examples. Section IV, we compare directly sixth order and fourth order theories. In Section V, the DSA formalism for theories of order eight is set up
together with other three examples. In Section VI, we conclude.

\section{Basic equations}\label{BasicEq}
The general action for a relativistic theory of gravity of order six is given by \cite{Gottlober:1989ww,Wands}
\beq
\label{ActionRBR}
{\mathcal A}=\int d^4x\sqrt{-g} \left[f(R,\Box R)+{\cal L}_m\right]\,.
\eeq
where $g$ is the metric determinant of the metric
$g_{\mu\nu}$, $f$ is a generic function of the Ricci scalar $R$ and of its d'Alembertian $\Box R$,  and  ${\cal L}_m$ is the standard matter Lagrangian. This theory is in general of order eight in the derivative of the metric. Since we consider the boundary terms as irrelevant, integrating by parts leads to a series  of relevant properties in the theory above. First, it is important to note that, not differently from the case of the Einstein-Hilbert action, if $f$ is linear in $\Box R$ the theory is only of order six. In fact, any non linear term in $\Box R$ appearing in $f$ can always be recast as a higher-order term. Thus, for example, $(\Box R)^2$  can be written as
\begin{equation}
(\Box R)^2= \nabla_\mu \left(\nabla^\mu R \Box R-R\nabla^\mu \Box R\right)+ R \Box^2 R\,.
\end{equation}
Thus, terms of the type $(\Box R)^n$ can be converted into terms of the
form $R\Box^n R$. In general, therefore, the class of theories of gravity with Lagrangian
\beq
\label{ActionRBR_Sim}
{\mathcal A}
=\int d^4x\sqrt{-g} \left[f_0(R)+ \sum_{i=1}^{n} a_i(R)\,(\Box R)^i+{\cal L}_m\right]\,,
\eeq
where the $a_i$ are  functions of the Ricci scalar, will have the same equations of motion of a theory whose action contains terms of higher order like, e.g., $R^n\Box^n R$. In this sense the analysis given in the following will extend also to this specific class of theories. 
In the following we will start  describing the general theory and then, using  the considerations above, we will present explicitly a dynamical systems formalism for Lagrangians of the type of Eq.~\eqref{ActionRBR_Sim}.

Variation of Eq.~\eqref{ActionRBR_Sim} upon the metric tensor gives the gravitational field equations
\begin{equation}
 \begin{split}
{\g} G_{\mu\nu}=&\frac{1}{2} g_{\mu\nu}[f-{\g}R]+{\g}\ddemunu-
\gd\Box {\g}\\
&-\frac{1}{2}\gd[{\fo}_{;\gamma}R^{;\gamma}+{\fo}\Box R]\\
&+{\fo}_{;(\mu}R_{;\nu)}+ T_{\mu\nu}\,,
 \end{split}
\end{equation}
where $T_{\mu\nu}$ is the standard stress energy tensor and 
\beq
\label{6.4}
{\g}=\frac{\pa f}{\pa R}+\Box{\fo}\,,\;\;\;\;\;
{\fo}=\frac{\pa f}{\pa \Box R}\,.
\eeq

We assume a Friedmann-Lema\^{\i}tre-Robertson-Walker metric
with an expansion factor denoted by $a(t)$ and
spatial curvature $k=-1,0.1$,
and further assume
the matter component to be an isotropic perfect fluid, i.e., $T_a^b=\left(-\mu,p,p,p\right)$,
where $\mu$ is the energy density and $p$ is the pressure of the fluid. 
The cosmological equations are then usually written as
\begin{equation}
 \begin{split}
{\g} \left(H^2+\frac{k}{a^2}\right)=&\frac{1}{6}\left(R {\g}-f+{\fo}\Box R+\dot{R}\dot{\fo}\right)\\
&-H\dot{\g}+\frac{\mu}{3}\;,
\label{CosmicEq}
\end{split}
\end{equation}
\begin{equation}
 \begin{split}
{\g} \left(\dot{H}+H^2\right)=&-\frac{1}{6}\left(f-R{\g}-{\fo}\Box R\right)-\frac{1}{3}\dot{R}\dot{\fo}-\\
&\frac{1}{2}\ddot{\g}-\frac{1}{2}H\dot{\g}-\frac{1}{6}(\mu+3 p)\;,
\label{CosmicEq2}
\end{split}
\end{equation}
where 
\begin{equation}
H=\frac{\dot a}{a},
\label{H1}
\end{equation}
is the Hubble parameter, a dot $\dot{\ }$ denotes a derivative with respect to time, and
\begin{equation}\label{RBRDef}
 \begin{split}
& R=6\left[\dot{H}+2H^2+\frac{k}{a^2}\right]\,,\\
& \Box R=-\ddot{R}-3H\dot{R}\,.
  \end{split}
\end{equation}
With an abuse of terminology, we will sometimes refer to Eq.~\rf{CosmicEq} as the  Friedmann equation and to Eq.~\rf{CosmicEq2} as the Raychaudhuri equation.

We introduce now  the logarithmic time
\begin{equation}
N=\ln\frac{a}{a_0}\,.
\label{time1}
\end{equation}
where $a_0$ is a constant with units of length that represents the value of the scale factor at the initial time $t=t_0$. We also define a set of seven
parameters as
\begin{equation} \label{HubbleVarDSN}
\begin{split}
&{\mathfrak q} =\frac{H^{(1)}_{N}}{H},\quad {\mathfrak j} =\frac{H^{(2)}_{N}}{H},\quad{\mathfrak s}=\frac{H^{(3)}_{N}}{H},\quad{\mathfrak s}_1=\frac{H^{(4)}_{N}}{H},\\
&{\mathfrak s}_2=\frac{H^{(5)}_{N}}{H},\quad{\mathfrak s}_3=\frac{H^{(6)}_{N}}{H}\quad{\mathfrak s}_4=\frac{H^{(7)}_{N}}{H}\,,
\end{split}
\end{equation}
where $H^{(i)}_{N}$ represent the $i$th-derivative of $H$ with respect to $N$.
One can write the above equations in terms of these variables, but  this is a long and rather tedious exercise which does not really add anything to the understanding of the problem. For this reason we will not show them here, giving directly the equations in terms of the dynamical variables in the following sections.
\section{Dynamical System Approach for the sixth-order case}

\subsection{The basic equations}\label{BasicEqb}

Let us start looking at the sixth-order case, i.e., $n=1$. Recalling the argument of the previous section, all theories of order six that have the form $f(R,\Box R)$ can be written without loss of generality as 
\begin{equation}
f= f_1(R)+ f_2(R) \Box R\,,
\label{fact1}
\end{equation}
where $f_1$ and $f_2$ are in general different functions of $R$.
Eq.~(\ref{fact1})
has the immediate consequence that ${\mathfrak s}_3$ and ${\mathfrak s}_4$ are not present  in the cosmological equations and the analysis of this classes of modes is greatly simplified.

In order to apply the scheme presented in Ref.~\cite{Carloni:2015jla} the action will need to be written in a dimensionless way. We introduce therefore the constant $R_0$, with $R_0>0$, which has dimension of the inverse of a length squared and we will consider the function $f$ of the type $f= R_0 \bar{f}(R_0^{-1} R, R_0^{-2}\Box R)$, for some function $\bar f$. This implies the definition of an auxiliary dynamical variable related to $R_0$.  
We then define the set of dynamical variables
\begin{align} \label{DynVar}
\begin{split}
&\mathbb{R}=\frac{R}{6 H^2},\quad \mathbb{B}=\frac{\Box R}{6H^4},\quad  \mathbb{K}=\frac{k}{a^2 H^2},\quad \Omega =\frac{\mu }{3H^2},\\ &\mathbb{J}={\mathfrak j},\quad \mathbb{Q}={\mathfrak q},\quad \mathbb{S}={\mathfrak s},\quad \mathbb{S}_1={\mathfrak s}_1
,\quad \mathbb{A}=\frac{R_0}{H^2}\,.
\end{split}
\end{align}
Note that in the above setting $\mathbb{A}$ and $\Omega $ are defined positive so that all fixed points with $\mathbb{A}<0$ or $\Omega<0$ should be excluded. The Jacobian of this variable definition reads
\begin{align} \label{Jac6}
M_6=-\frac{1}{108 a^2 H^{32}},
\end{align}
which implies that the variables are always regular if $H\neq 0$ and $a \neq 0$.

The requirement to have a closed system of equations
demands the introduction of the auxiliary quantities
\begin{align}\label{XYZT1}
\begin{split}
& {\bf X}_1\left(\mathbb{A},\mathbb{R}\right)= \frac{f_1\left(\mathbb{A},\mathbb{R}\right)}{H^2},\\ 
& {\bf X}_2\left(\mathbb{A},\mathbb{R} \right)= H^2f_2\left(\mathbb{A},\mathbb{R}\right),\\ 
&{\bf Y}_1\left(\mathbb{A},\mathbb{R} \right)= f'_{1}\left(\mathbb{A},\mathbb{R}\right),\\ 
& {\bf Y}_2\left(\mathbb{A},\mathbb{R} \right)=H^4 f'_{2}\left(\mathbb{A},\mathbb{R}\right),\\
& {\bf Z}_1\left(\mathbb{A},\mathbb{R}\right)= H^2 f''_1\left(\mathbb{A},\mathbb{R}\right),\\ 
& {\bf Z}_2\left(\mathbb{A},\mathbb{R} \right)=H^6 f''_2\left(\mathbb{A},\mathbb{R}\right),\\ 
& {\bf W}_1\left(\mathbb{A},\mathbb{R}\right)=H^4 f^{(3)}_1\left(\mathbb{A},\mathbb{R}\right),\\ 
&{\bf W}_2\left(\mathbb{A},\mathbb{R} \right)=H^8 f^{(3)}_2\left(\mathbb{A},\mathbb{R}\right),\\ 
&{\bf T}\left(\mathbb{A},\mathbb{R} \right)= H^{10}f^{(4)}_{2}\left(\mathbb{A},\mathbb{R}\right),\\
\end{split}
\end{align}
where the prime represents the derivative with respect to the Ricci scalar
$\mathbb{R}$. The dynamical equations can be written as

\begin{align}\label{DynSys6}
\begin{split}
\DerN{\mathbb{R}}&=\mathbb{J}+(\mathbb{K}-2) \mathbb{K}-(\mathbb{R}-2)^2 ,\\ 
\DerN{\mathbb{B}}&=\mathbb{B} (3 \mathbb{K}-3 \mathbb{R}+7)+\frac{1}{2}
   \left[\mathbb{J}+\mathbb{K}^2-2 \mathbb{K} (\mathbb{R}+1)+\mathbb{R}^2-4\right]^2\\
   &~~~+\frac{1}{\mathbf{Y}_2}\left\{\frac{\Omega }{12}+18 \left[\mathbb{J}+\mathbb{K}^2-2
   \mathbb{K} (\mathbb{R}+1)+\mathbb{R}^2-4\right]^3 \mathbf{W}_2\right.\\
   &~~~-\frac{1}{72} \mathbf{X}_1-\frac{1}{12} (\mathbb{K}-\mathbb{R}+1)
   \mathbf{Y}_1+\left(2-\frac{\mathbb{J}}{2}-\frac{\mathbb{K}^2}{2}+\mathbb{K}
   \mathbb{R}+\mathbb{K}-\frac{\mathbb{R}^2}{2}\right)
    \mathbf{Z}_1\\
   &~~~-3 \left[4 \mathbb{B}-(\mathbb{K}-\mathbb{R}-5)
   \left(\mathbb{J}+\mathbb{K}^2-2 \mathbb{K}
   (\mathbb{R}+1)+\mathbb{R}^2-4\right)\right] \times\\
   &~~~\left.+\left[\mathbb{J}+\mathbb{K}^2-2 \mathbb{K} (\mathbb{R}+1)+\mathbb{R}^2-4\right]
   \mathbf{Z}_2 \right\},\\
\DerN{\Omega} &=\Omega  (1-3w+2 \mathbb{K}-2 \mathbb{R}),\\ 
\DerN{\mathbb J}&=\mathbb{J} (5 \mathbb{K}-5
   \mathbb{R}+3)+(\mathbb{K}-\mathbb{R}) \left[\mathbb{K}^2-\mathbb{K} (2 \mathbb{R}+7)+\mathbb{R}
   (\mathbb{R}+5)\right],\\
   &~~~-\mathbb{B}-22 \mathbb{K}+20 \mathbb{R}-12,\\
\DerN{\mathbb{K}}&=2 \mathbb{K} (\mathbb{K}-\mathbb{R}+1),\\
\DerN{\mathbb{A}}&=2 \mathbb{A}(\mathbb{K}-\mathbb{R}+2)\,.
\end{split}
\end{align}
To eliminate the equations for $\S_1$, $\Q$, $\S$ 
we have implemented in the equations above
the Friedmann equation, Eq.~\rf{CosmicEq},
and the following constraints coming from the definition  of $R$ and $\Box R$ in Eqs.~\eqref{RBRDef}:
\begin{align}\label{ConstrRBR}
\begin{split}
\mathbb{R}=&\mathbb{K}+\mathbb{Q}+2,\\
\mathbb{B}=&-4 \mathbb{J} \mathbb{Q}-7 \mathbb{J}+2 \mathbb{K} \mathbb{Q}+2\,
   \mathbb{K}\\ 
&-\mathbb{Q}^3-11 \mathbb{Q}^2-12 \mathbb{Q}-\mathbb{S}\,.
 \end{split}
\end{align}
As mentioned in the Introduction, for the sake of simplicity, we do not report here the full cosmological equations in terms of the variables in Eq.~\eqref{HubbleVarDSN}. They are very long and  their full form does not add much to the understanding of the derivation of the  fixed points and their properties. The reader can find some examples of the full form of these equations in the Appendix \ref{App}. 

The solutions associated to the fixed points can be derived  writing Raychaudhuri equation, Eq.~\rf{CosmicEq2}, in terms of the variables
given in Eq.~\eqref{HubbleVarDSN} and solving for $\mathfrak{s}_2$. Since Eq.~\rf{CosmicEq2} is linear in $\mathfrak{s}_2$ via the term $\ddot{\g}$, this does not present any problem. From the definition of $\mathfrak{s}_2$, in a fixed point we can write
\begin{align}\label{RAy6Ord}
\begin{split}
& \frac{1}{H}\frac{ d^5{H}}{d N^5}=\mathfrak{s}^*_2,
\end{split}
\end{align}
where here, and in the following, the asterisk $*$ indicates the value of a variable in a fixed point. The characteristic polynomial of Eq.~\rf{RAy6Ord} has one real and two pairs of complex roots. Hence, we can write an exact solution for $H(N)$:
\begin{equation} \label{SolHFixPointsGen6}
\begin{split}
H=\sum_{i=0}^{2}\exp \left(p\,\alpha_i N\right)\left[H_i \cos\left(\beta_i p N\right)+\bar{H}_i \sin\left(\beta_i p N\right)\right],\\
\end{split}
\end{equation}
where $p=-\sqrt[5]{\mathfrak{s}^*_2}$, $H_i$ and $\bar{H}_i$ are integration constants and  $a_i$ and $b_i$ are given by
\begin{equation}
\begin{array}{ll}
\alpha_0=-1, & \beta_0=0,\\
\alpha_1=\frac{1}{4}(\sqrt{5}+1), &\beta_1=\sqrt{\frac{5}{8}-\frac{\sqrt{5}}{8}},\\
\alpha_2=\frac{1}{4}(1-\sqrt{5}), &\beta_2=\sqrt{\frac{5}{8}+\frac{\sqrt{5}}{8}},
\end{array}
\end{equation}
i.e. are connected with the fifth root of unity. We are obviously interested in real solutions, which can be derived by a suitable redefinition of the integration constants.  The solutions above are oscillating, however  they do not correspond to oscillating scale factors. Indeed the scale factor is given by the equation
\begin{equation} \label{EqaFixPoints6}
\dot{a}=\sum_{i=0}^{2}a^{1+p\,\alpha_i}\left[H_i \cos\left(\beta_i p \ln a\right)+\bar{H}_i \sin\left(\beta_i p \ln a\right)\right],\\
\end{equation} 
which can be solved numerically. Notice that this solution, like Eq.~\eqref{SolHFixPointsGen6} is parameterised only by the quantity $p$ and therefore $\mathfrak{s}^*_2$. In the following we will characterise these solutions only by the value of 
$\mathfrak{s}^*_2$.

\begin{figure}[h] 
\includegraphics[scale=0.6]{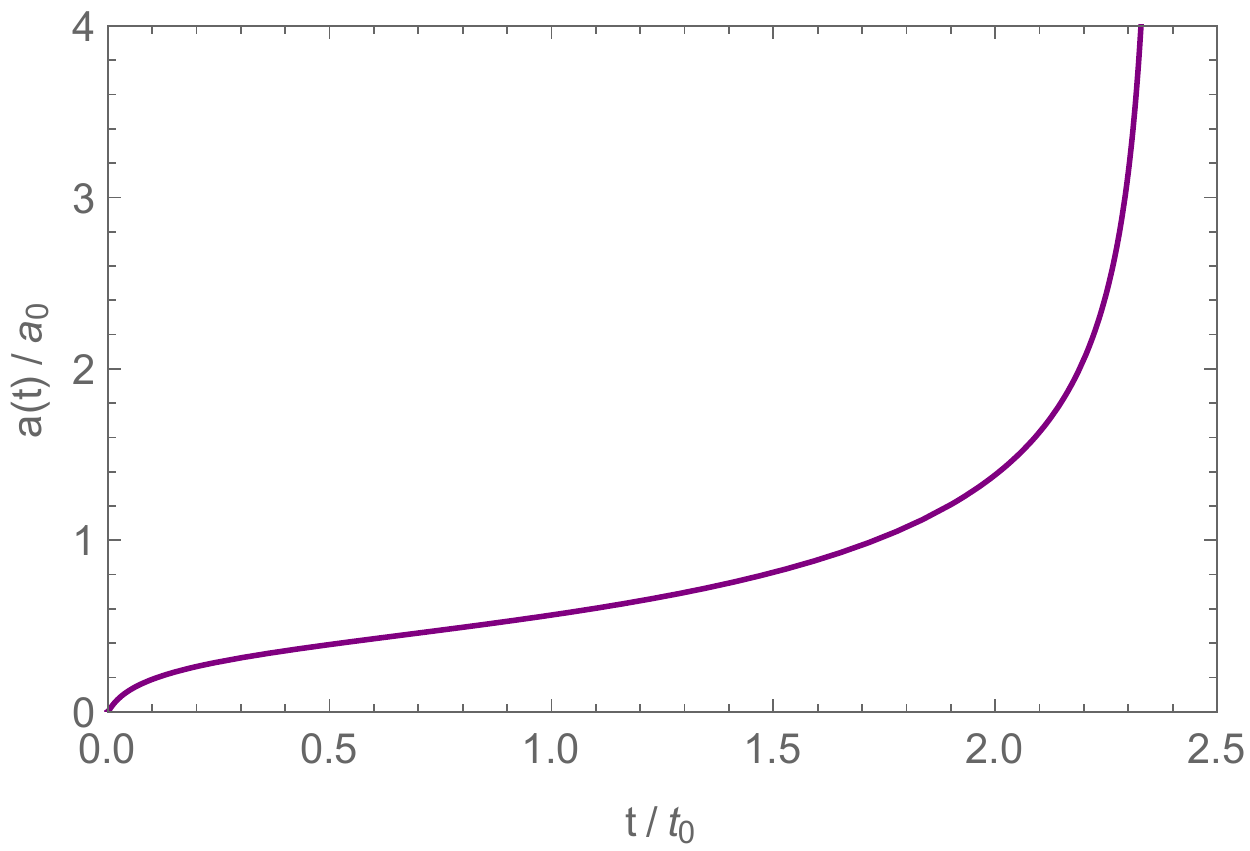}
\caption{Behaviour of the scale factor in a fixed point $\mathfrak{s}^*_2=0$. All integration constants have been chosen to be one.}
\label{Fig:POINT_C_6TH_LIN}
\end{figure}

In the case $\mathfrak{s}^*_2=0$ the equation to solve is 
\begin{equation}\label{SolH}
\dot{N}=\sum_{i=0}^{4}H_i N^i,
\end{equation}
which in terms of $a$ reads
\begin{equation}\label{SolPF}
\dot{a}=a\sum_{i=0}^{4}H_i (\ln a)^i.
\end{equation}
Eq.~\eqref{SolH} can be solved by separation of variables and it has a solution that depends on the roots of the polynomial in $N$ on the right hand side. In particular, the scale factor can have a finite time singularity if any of the roots of the polynomial are complex, otherwise it evolves asymptotically towards a constant value of the scale factor, i.e., a static universe.  Therefore,  a fixed point with  $\mathfrak{s}_2=0$, will correspond to one of these two cosmic histories depending on the value of the constants $H_i$. Considering that the solution given in Eq.~\rf{SolPF} can be viewed as an approximation of the general integral of the cosmology, then the values of the constants $H_i$ should match the initial conditions of the orbit. This implies that the solution in the fixed point will depend on the initial condition of the orbit that reaches it. In Fig.~\ref{Fig:POINT_C_6TH_LIN} we show time dependence of the scale factor corresponding to this point. 

In the following we will examine two specific examples. The first one will show the phase space of a theory in which only sixth-order terms are present other than the Einstein-Hilbert one. This example will clarify the action of these terms. The second one will contain also fourth-order terms, so that the interaction between sixth  and fourth-order corrections can be observed explicitly. 

\subsection{Two Examples}\label{twocases}

\subsubsection{Case $f= R+ \gamma R \Box R$}\label{Example6_1}

In this case only the Einstein-Hilbert plus
sixth-order terms are present in the theory. It is an interesting example as
it clarifies the interplay between these terms.
In this case the action can be written as 
\begin{equation}\label{ActionGR+6Ord}
{\mathcal A}=\int d^4x\sqrt{-g} \left[  R+ \gamma R_0^{-2} R \Box R+{\cal L}_m\right]\,,
\end{equation}
which implies  $f_1=R_0^{-1} R$ and $f_2= \gamma R_0^{-3} R$. Then the only non-zero auxiliary quantities in Eq.~\rf{XYZT1} are
\begin{align}\label{XYZT1new}
\begin{split}
& {\bf X}_1\left(\mathbb{A},\mathbb{R}\right)= 6 \mathbb{R},\qquad {\bf X}_2\left(\mathbb{A},\mathbb{R} \right)= \frac{6\gamma \mathbb{R}}{\mathbb{A}},\\ 
&{\bf Y}_1\left(\mathbb{A},\mathbb{R} \right)= 1,\qquad {\bf Y}_2\left(\mathbb{A},\mathbb{R} \right)= \frac{\gamma}{\mathbb{A}^2},
\end{split}
\end{align}
and the  Friedmann and Raychaudhuri equations,   Eqs.~\rf{CosmicEq} and~\rf{CosmicEq2} respectively,
can be found  in Eq.~(\ref{a1a1}) of
Appendix~\ref{App}.

\begin{figure}[!ht]
     \subfloat[Point $\mathcal A$\label{P_A_Ex1}]{%
       \includegraphics[width=0.45\textwidth]{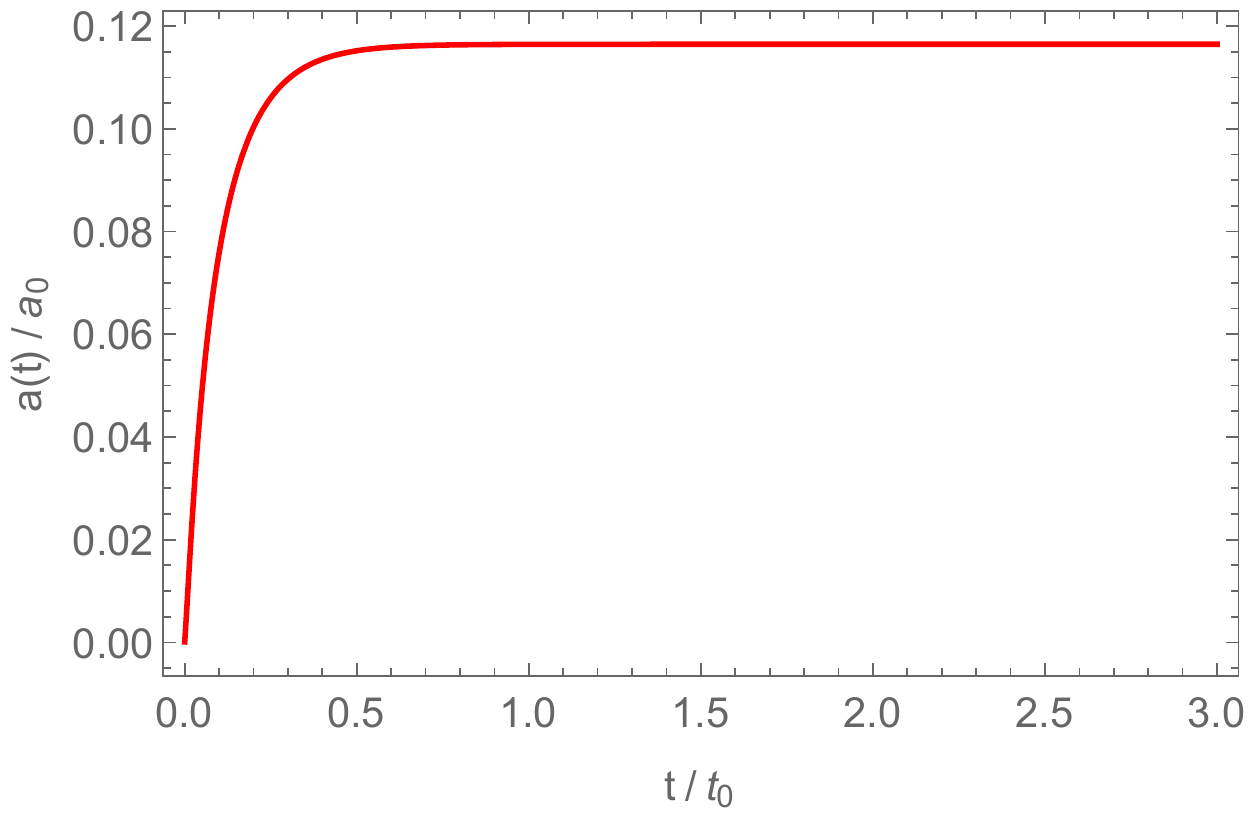}
     }
     \hfill
     \subfloat[Point $\mathcal B$\label{P_B_Ex1}]{%
       \includegraphics[width=0.45\textwidth]{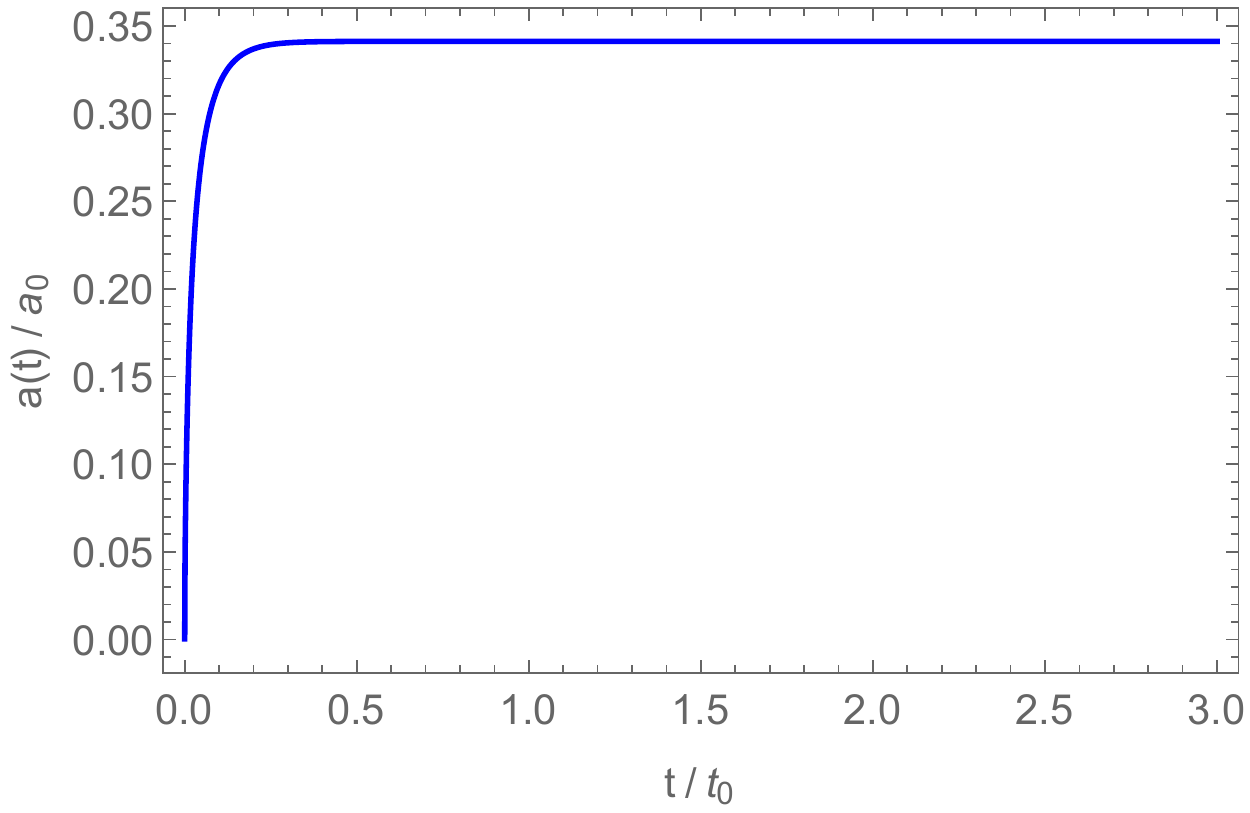}
     } \\
     \subfloat[Point ${\mathcal{I}}_1$\label{P_D_Ex1}]{%
       \includegraphics[width=0.45\textwidth]{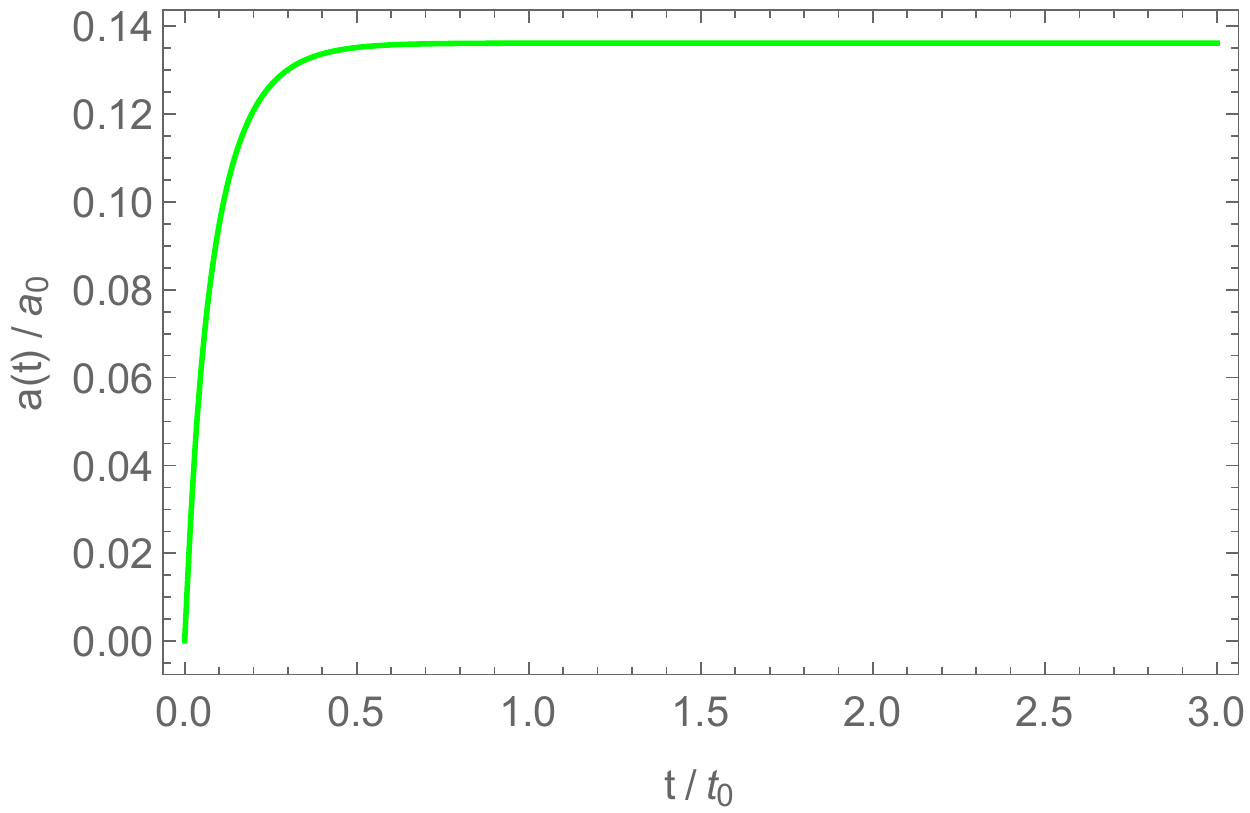}
     }\hfill
     \subfloat[Point ${\mathcal{I}}_2$\label{P_E_Ex1}]{%
       \includegraphics[width=0.45\textwidth]{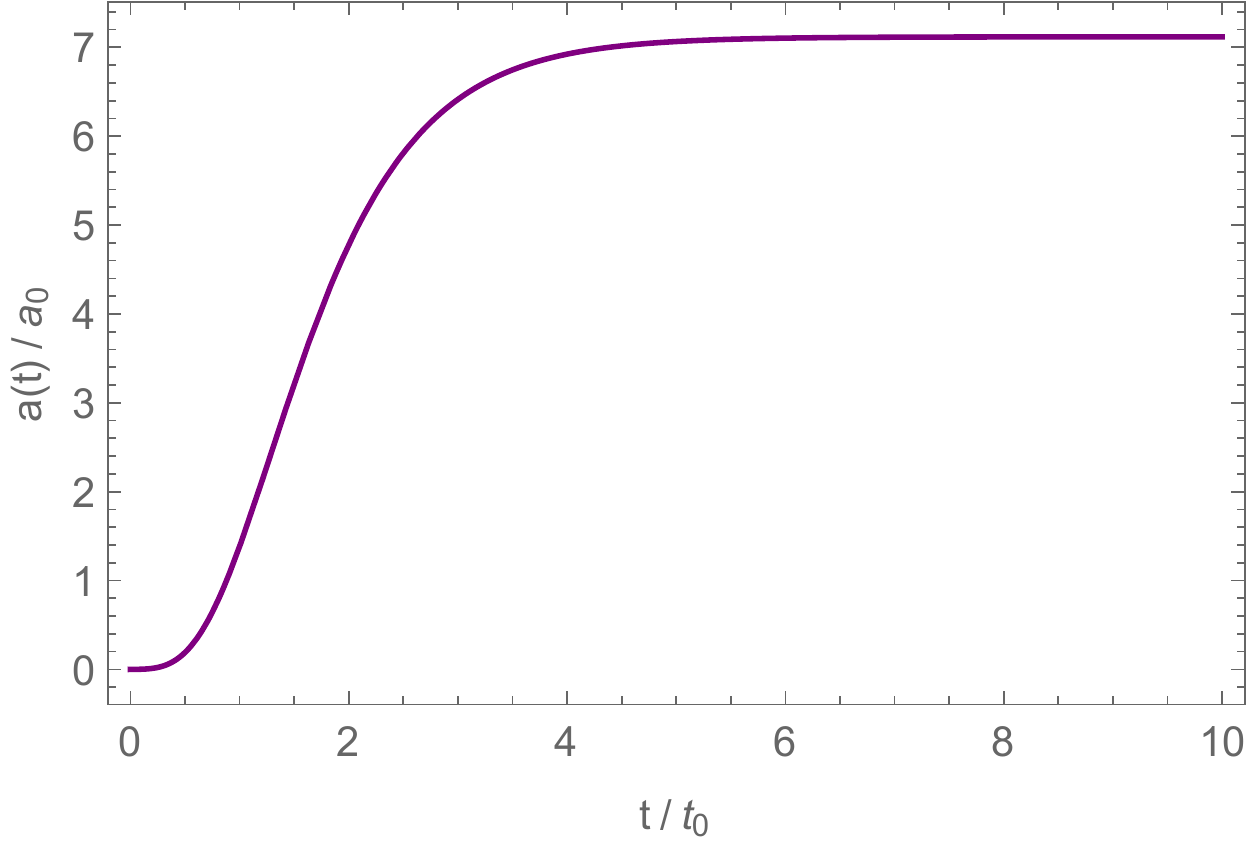}
     }
     \caption{Behaviour of the scale factor in the fixed points of the phase space of the theory $f(R,\Box R)=R_0^{-1} R+ R_0^{-3} R \Box R$. The integration constants have  all been chosen to be one.}
     \label{fig_ex6-1}
   \end{figure}

The  dynamical system in Eq.~\rf{DynSys6} becomes
\begin{align}\label{DynSysRRBR}
\begin{split}
\DerN{\mathbb{R}}=&\mathbb{J}+(\mathbb{K}-2) \mathbb{K}-(\mathbb{R}-2)^2 ,\\ 
\DerN{\mathbb{B}}=&\mathbb{B} (3 \mathbb{K}-3 \mathbb{R}+7)-\frac{\mathbb{A}^2 (\mathbb{K}-\Omega +1)}{12 \gamma }
\\&+\frac{1}{2} \left(\mathbb{J}+\mathbb{K}^2-2 \mathbb{K} (\mathbb{R}+1)+\mathbb{R}^2-4\right)^2,\\
\DerN{\mathbb J}=&\mathbb{J} [5 (\mathbb{K}-
   \mathbb{R})+3]-\mathbb{B}-22 \mathbb{K}+20 \mathbb{R}-12
\\&+(\mathbb{K}-\mathbb{R}) \left[\mathbb{K}^2-\mathbb{K} (2 \mathbb{R}+7)+\mathbb{R}
   (\mathbb{R}+5)\right],\\
\DerN{\Omega} =&\Omega  (1-3w+2 \mathbb{K}-2 \mathbb{R}),\\ 
\DerN{\mathbb{K}}=&2 \mathbb{K} (\mathbb{K}-\mathbb{R}+1),\\
\DerN{\mathbb{A}}=&2 \mathbb{A}(\mathbb{K}-\mathbb{R}+2)\,.
\end{split}
\end{align}
The system presents three invariant submanifolds $\Omega=0$, $\K=0$ and $\A=0$, therefore only points that belong to all of these three submanifolds can be  true global attractors. The fixed points of the system can be found in Table~\ref{TavolaRBR}, together with their associated solutions which are represented graphically in Fig.~\ref{fig_ex6-1}. Point $\mathcal{C}$ has a solution of the type described by Eq.~\eqref{SolPF} and as such can indicate the occurrence of a finite time singularity.
\begin{table}[h]
\samepage
\centering
\caption{Fixed points of $f(R,\Box R)=R_0^{-1} R+ R_0^{-3} R \Box R$ and the parameter ${\mathfrak s}_2$ that characterise its solution. Here A stays for attractor, S for saddle, NHR for non hyperbolic repeller, NHS for non hyperbolic saddle. } \label{TavolaRBR}
\begin{tabular}{lllcc} \hline\hline
Point & Coordinates  & Solution & Stability \\
&$\{\mathbb{R},\mathbb{B},\mathbb{J},\Omega,\mathbb{K}, \A\}$ &  \\ \hline\\
$\mathcal{A}$ & $\left\{ 0,0,1,0, -1 , 0\right\}$  & ${\mathfrak s}_2=-1$ & NHS\\ \\
$\mathcal{B}$ & $\left\{ 0, 0, 4, 0, 0, 0\right\}$  & ${\mathfrak s}_2=-32$ & \makecell[c]{ NHR for $w < 1/3$ \\ NHS for $w > 1/3$}\\ \\
$\mathcal{C}$ & $\left\{ 2, 0, 0,0, 0, 0\right\}$  & ${\mathfrak s}_2=0 \rightarrow$ \rf{SolPF}& NHS \\ \\
$\mathcal{I}_1$ & $\left\{ a_\mathcal{I}^-, b_\mathcal{I}^-,c_\mathcal{I}^+,0,0,0\right\}$  & ${\mathfrak s}_2={\mathfrak s}_{2}^{\mathcal{I}_1}$ & S  \\ \\
$\mathcal{I}_2$ & $\left\{ a_\mathcal{I}^+, b_\mathcal{I}^+,c_\mathcal{I}^-,0,0,0\right\}$  & ${\mathfrak s}_2={\mathfrak s}_2^{\mathcal{I}_2}$  & A  \\ \\
\hline\\
\multicolumn{5}{c}{\makecell[c]{$a_\mathcal{I}^\pm=\frac{1}{10}\left(16\pm\sqrt{46}\right) $\qquad $b_\mathcal{I}^\pm=-\frac{9}{250}\left(74\pm 9\sqrt{46}\right) $ \qquad $c_\mathcal{I}^\pm= \frac{1}{50}\left(31\pm4\sqrt{46}\right)$\\ \\ ${\mathfrak s}_{2}^{\mathcal{I}_1} =-\frac{1}{10}\left(4+\sqrt{46}\right)$\qquad ${\mathfrak s}_2^{\mathcal{I}_2}=-\frac{1}{10}\left(4-\sqrt{46}\right)$
 }}\\\\\hline\hline\\
 \end{tabular}
\end{table}

The stability of fixed points $\mathcal{B}$ for $w\neq 1/3$, $\mathcal{I}_1$ and $\mathcal{I}_2$,  can be deduced by the Hartmann-Grobmann theorem and it is also shown in Table \ref{TavolaRBR}. Points $\mathcal{B}$ and $\mathcal{I}_1$ are unstable, but $\mathcal{I}_2$ is an attractor. Indeed this point is a global attractor for the cosmology as it lays on the intersection of the 
three invariant submanifolds of the phase space. The remaining points
$\mathcal{A}$, $\mathcal{B}$ for $w = 1/3$, and $\mathcal{C}$, are non hyperbolic, as they have a zero eigenvalue. Their stability can be analysed via the central manifold theorem \cite{Wiggins}.

For point $\mathcal{A}$, for example, defining the variables
\begin{align}\label{CMAex1}
\begin{split}
&x_1=12 \mathbb{B} (3 w+1)+\Omega,\\
&x_2=\frac{\mathbb{B}}{6}+\mathbb{J}+\frac{\Omega }{72 (3
   w-5)}-1,\\
&x_3= \mathbb{A},\\
&x_4=\frac{\mathbb{B}}{2}+\mathbb{J}-4 (\mathbb{K}+1)+4 \mathbb{R}+\frac{\Omega
   }{24 (3 w-1)}-1,\\
&x_5=\Omega,\\
&y=\frac{\mathbb{B}}{4}+\mathbb{J}-4 (\mathbb{K}+1)+2
   \mathbb{R}+\frac{\Omega }{144 (w-1)}-1,\\
\end{split}
\end{align}
 and  expressing the dynamical equations in the new variables, the equation of the center manifold $\mathbf{x}=\mathbf{h}(y) $ is given by  the system of equations
\begin{equation}\label{CM_Eq}
\DerN{\mathbf{x}}= \frac{d\mathbf{h}(y)}{d y}\DerN{y},
\end{equation}
where the vector  $\mathbf{x}$ has components $\mathbf{x}=\{x_1,x_2,x_3,x_4,x_5\}$. Solving the above system per series at third-order, i.e., setting
\begin{equation}
\mathbf{x}=\sum_{i=2}^{3} \mathbf{a}_i y^i +O(y^4),
\end{equation}
gives the solutions
\begin{equation}
\begin{split}
& \mathbf{a}_2=\left\{ -\frac{3}{2} (3 \omega +1), \frac{1}{24},0,-\frac{1}{8},0 \right\}, \\
& \mathbf{a}_3=\left\{-\frac{3}{16} (3 \omega +1),\frac{5}{384},0,-\frac{5}{128},0 \right\}.
\end{split}
\end{equation}
Note that the center manifold coincides with the center space for the variables $x_3$ and $x_5$. The equation for the central manifold is
 \begin{equation}
\DerN{y}= \frac{1}{8}y^2+ O(y^3).
\end{equation}
Using the Shoshitaishvili theorem we can conclude that  the stability of $\mathcal{A}$ is a complex combination of saddle nodes in each planes $(x_i, y)$  with $i\neq3,5$ and the center spaces for  $x_3$ and $x_5$. Looking at the coefficients of $\mathbf{a}_2$ we can conclude that this point is in general unstable. 

We can  apply the same procedure to the other non-hyperbolic points. However, we can also evaluate the character of these points in a faster way. In fact, point $\mathcal{A}$ has eigenvalues $\{4,-2,2,2,0,-(1+3w)\}$, i.e., with alternate signs. Therefore, regardless of the behaviour of the central manifold, this point is in fact always a saddle. This implies that in some cases we can evaluate the stability of a non hyperbolic fixed point without analysing in detail the central manifold. Clearly this is insufficient if the aim is to characterise the exact behaviour of the flow in the phase space. However, since we are mainly interested in the attractors in the phase space, such less precise analysis will be sufficient here.

\subsubsection{Case $f= R+ \alpha R^3 +\gamma R \Box R$}\label{Example6_2}

In this case the Einstein-Hilbert plus fourth- and 
sixth-order correction terms are present in the theory and 
the
interaction between them can be appreciated.
Consider then the  action
\begin{equation}
{\mathcal A}=\int d^4x\sqrt{-g} \left[ R+ \alpha  R_0^{-2} R^{3}+ \gamma R_0^{-2} R \Box R+{\cal L}_m\right]\,,
\end{equation}
which implies  $f_1=R_0 R+\alpha  R_0^3 R^{3}$ and $f_2= R_0^{-3} R$. Hence the only non zero auxiliary quantities in Eq.~\rf{XYZT1} are
\begin{align}\label{XYZT1newn}
\begin{split}
& {\bf X}_1\left(\mathbb{A},\mathbb{R}\right)= 6\mathbb{R}+\frac{216\alpha \mathbb{R}^3}{\mathbb{A}^2},\qquad {\bf X}_2\left(\mathbb{A},\mathbb{R} \right)=\frac{6\gamma \mathbb{R}}{\mathbb{A}^2},\\ 
&{\bf Y}_1\left(\mathbb{A},\mathbb{R} \right)= 1+\frac{108\alpha \mathbb{R}^2}{\mathbb{A}^2}, \qquad {\bf Y}_2\left(\mathbb{A},\mathbb{R} \right)=\frac{\gamma}{\mathbb{A}^2}, \\
&{\bf Z}_1\left(\mathbb{A},\mathbb{R} \right) =\frac{36 \alpha  \mathbb{R}}{\mathbb{A}^2},\qquad {\bf W}_1\left(\mathbb{A},\mathbb{R} \right) =\frac{6 \alpha }{\mathbb{A}^2},
\end{split}
\end{align}
and the cosmological equations can be decoupled to give an explicit equation for
$\mathbb{S}_1$ and $\mathbb{S}_2$. These are given 
in Eq.~(\ref{b1b1}) of
Appendix~\ref{App}.

The  dynamical system Eq.~\rf{DynSys6} becomes
\begin{align}\label{DynSysRR3RBR}
\begin{split}
\DerN{\mathbb{R}}&=\mathbb{J}+(\mathbb{K}-2) \mathbb{K}-(\mathbb{R}-2)^2 ,\\ 
\DerN{\mathbb{B}}&=\mathbb{B} (3 \mathbb{K}-3 \mathbb{R}+7)+\frac{1}{2} \left(\mathbb{J}+\mathbb{K}^2-2 \mathbb{K}
   (\mathbb{R}+1)+\mathbb{R}^2-4\right)^2\\
&~~~-\frac{3 \alpha  }{\gamma }  \mathbb{R} \left[6(\mathbb{J}+(\mathbb{K}-2)
   \mathbb{K}-4)+3 \mathbb{R} (1-3\mathbb{K} +1)+4 \mathbb{R}^2\right]+\frac{\A^2}{12 \gamma } (-\mathbb{K}+\Omega -1),\\
\DerN{\Omega} &=\Omega  (1-3w+2 \mathbb{K}-2 \mathbb{R}),\\ 
 \DerN{\mathbb J} &=-\mathbb{B}+\mathbb{J} (5
   \mathbb{K}-5 \mathbb{R}+3)+(\mathbb{K}-\mathbb{R}) \left(\mathbb{K}^2-\mathbb{K} (2 \mathbb{R}+7)+\mathbb{R}
   (\mathbb{R}+5)\right)-22 \mathbb{K}+20 \mathbb{R}-12,\\
\DerN{\mathbb{K}}&=2 \mathbb{K} (\mathbb{K}-\mathbb{R}+1),\\
\DerN{\mathbb{A}}&=2 \mathbb{A}(\mathbb{K}-\mathbb{R}+2)\,.
\end{split}
\end{align}
The system above presents the same invariant submanifolds of Eq.~\rf{DynSysRRBR} and therefore we can draw the same conclusions for the existence of global attractors. Table~\ref{TavolaRR3RBR} summarises the fixed points for this system with the associated solution and their stability. All  the solutions associated to the fixed points are characterised by $\mathfrak{s}_2\neq0$ with the exception of $\mathcal{C}$ which is characterised by the solution Eq.~\rf{SolPF}. 

\begin{table}[htb]
\begin{center}
\caption{Fixed points of $f(R,\Box R)=R_0^{-1} R+ R_0^{-3} R^3+ R_0^{-3} R \Box R$ and their  associated solutions. Here  A stays for attractor, R for repeller, S for saddle, NHS for non hyperbolic saddle
. The quantities $\mathbb{R}^*_i$ are the solutions ofEq.~\rf{EqPtH}.} \label{TavolaRR3RBR}
\begin{tabular}{llllll} \hline\hline
Point & Coordinates  & Solution & Existence/ &  Stability \\
&$\{\mathbb{R},\mathbb{B},\mathbb{J},\Omega,\mathbb{K}, \A\}$ & parameter $\mathfrak{s}_2$  & Phsyical \\ \hline\\
$\mathcal{A}$ & $\left\{ 0,0,1,0, -1 , 0\right\}$  & $\mathfrak{s}_2=-1$ & always & NHS\\ \\
$\mathcal{B}$ & $\left\{ 0, 0, 4, 0, 0, 0\right\}$  & $\mathfrak{s}_2=-32$ & always & \makecell[l]{ R for $w < 1/3$ \\ S for $w > 1/3$}\\ \\
$\mathcal{C}$ & $\left\{ 2, 0, 0,0, 0, 12\sqrt{\alpha}\right\}$ & $\mathfrak{s}_2=0$ &$\alpha>0$&S\\ \\
$\mathcal{G}$ & $\left\{ 12+ \frac{2\gamma}{3\alpha}, 0, 1,0, 11+ \frac{2\gamma}{3\alpha}, 0\right\}$  & $\mathfrak{s}_2=\mathfrak{s}_\mathcal{G}$ &$\{\alpha,\gamma\}\neq0$ & S  \\ \\
$\mathcal{H}_1$ & $\left\{\mathbb{R}^*_1 , -6\mathbb{R}^*_1(\mathbb{R}^*_1-1)(\mathbb{R}^*_1-2),(\mathbb{R}^*_1-2)^2,0,0,0\right\}$  & $\mathfrak{s}_2=\sigma_{1}$  & Fig.~\ref{Fig:existPointH} &   Fig.~\ref{Fig:stabR1}  \\ \\
$\mathcal{H}_2$ & $\left\{\mathbb{R}^*_2 , -6\mathbb{R}^*_2(\mathbb{R}^*_2-1)(\mathbb{R}^*_2-2),(\mathbb{R}^*_2-2)^2,0,0,0\right\}$  & $\mathfrak{s}_2=\sigma_{2}$  & Fig.~\ref{Fig:existPointH} &   Fig.~\ref{Fig:stabR2}  \\ \\
$\mathcal{H}_3$ & $\left\{\mathbb{R}^*_3 , -6\mathbb{R}^*_3(\mathbb{R}^*_3-1)(\mathbb{R}^*_3-2),(\mathbb{R}^*_3-2)^2,0,0,0\right\}$  & $\mathfrak{s}_2=\sigma_{3}$  & Fig.~\ref{Fig:existPointH} &   S  \\ \\
\hline \\ 
\multicolumn{5}{l}{$\mathfrak{s}_\mathcal{G}=-577-5184\frac{\alpha}{\gamma}-11\frac{\gamma}{\alpha}$} \\ \\
\multicolumn{5}{l}{\makecell[l]{$\sigma_i= \frac{\alpha}{\gamma}\left(-150 \mathbb{R}_{*,i}^4+435 \mathbb{R}_{*,i}^3-252 \mathbb{R}_{*,i}^2\right)+101
   \mathbb{R}_{*,i}^5-610 \mathbb{R}_{*,i}^4+1306 \mathbb{R}_{*,i}^3-1180 \mathbb{R}_{*,i}^2+416 \mathbb{R}_{*,i}-32\neq 0$}}\\ \\
\hline\hline\\
 \end{tabular}
   \end{center}
\end{table}

\begin{figure}[p]   
\includegraphics[scale=0.55]{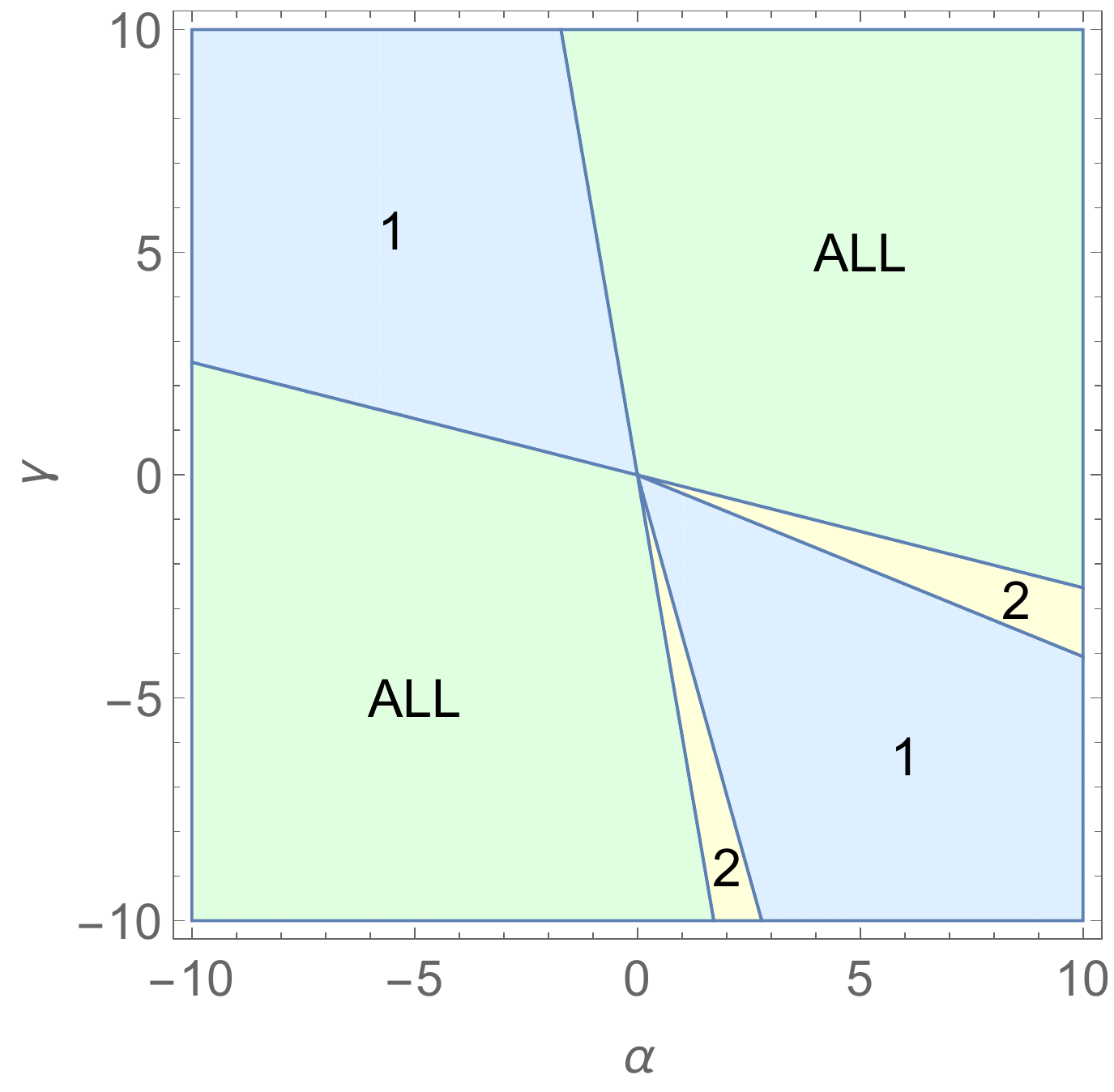}
\vskip -0.3cm
\caption{Region of the parameter space of $\alpha$ and $\gamma$ for which  the fixed points $\mathcal H$ exist. The number in the coloured area refer to the index $i$ of the point $\mathcal H_i$.}
\label{Fig:existPointH}
\includegraphics[scale=0.6]{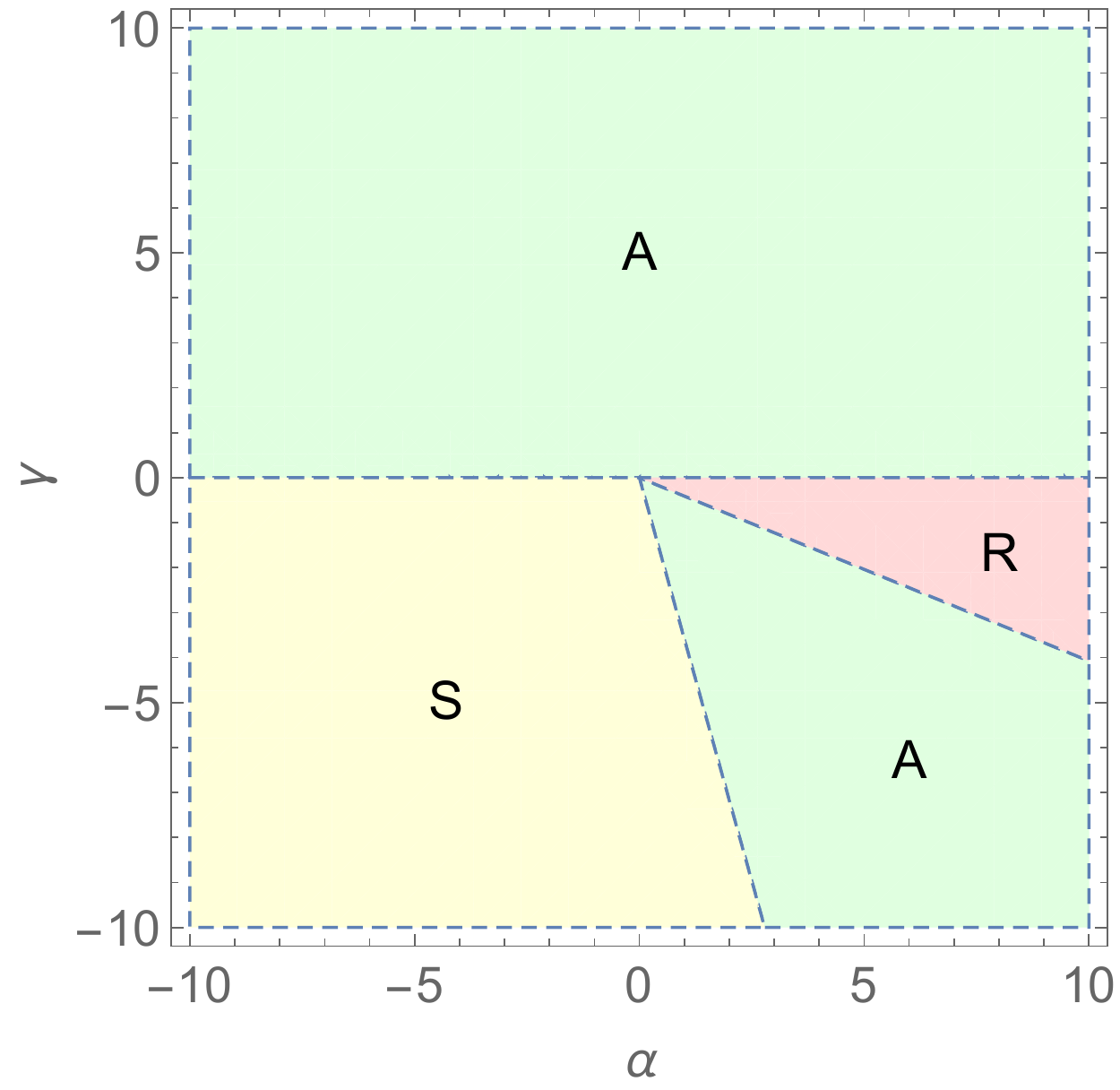}
\vskip -0.3cm
\caption{Stability of the fixed point $\mathcal{H}_1$. A stands for attractor (green), R stands for repeller (red), and S stands for saddle (yellow).}
\label{Fig:stabR1}
\includegraphics[scale=0.55]{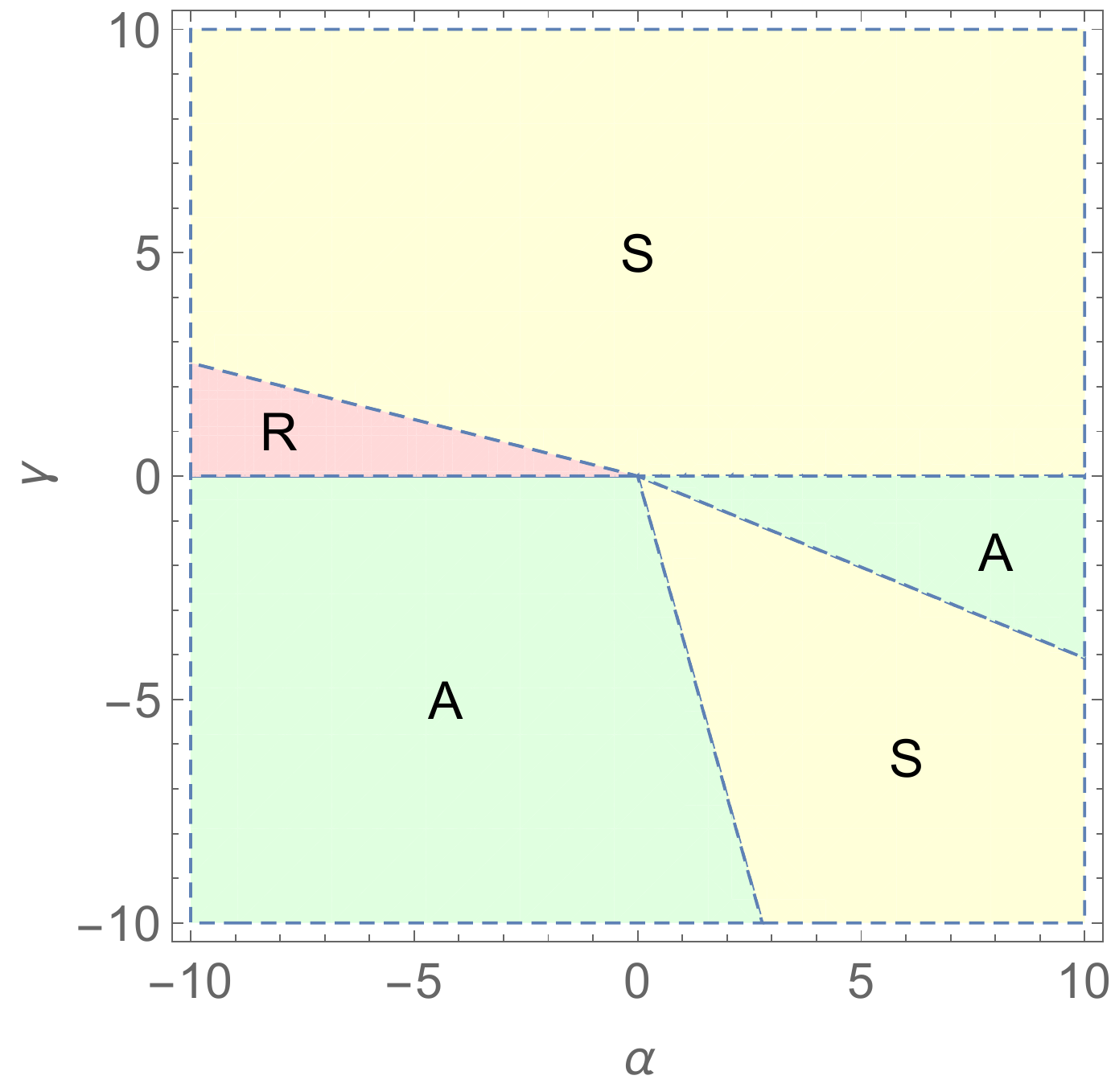}
\vskip -0.3cm
\caption{Stability of the fixed point $\mathcal{H}_2$. A stands for attractor (green), R stands for repeller (red), and S stands for saddle (yellow).}
\label{Fig:stabR2}
\end{figure}

Some of the fixed points exist only for specific values of the parameters $\alpha$ and $\gamma$. For example, the existence of $\mathcal{C}$ requires $\alpha>0$ and more complex conditions hold for the points $\mathcal{H}_i$ whose coordinates are determined by the equation
\begin{equation}\label{EqPtH}
3 \alpha  (21-10 \mathbb{R}^*_i) \mathbb{R}^*_i+2 \gamma  (\mathbb{R}^*_i-2) [2\mathbb{R}^*_i (5 \mathbb{R}^*_i-16)+21]=0\,.
\end{equation}
In Fig.~\ref{Fig:existPointH} we plot the region of existence of these points.
With the exception of point $\mathcal{A}$ all the other fixed points are hyperbolic, although their stability depends on the parameters $\alpha$ and $\gamma$. This complex dependence makes very complicated to make general statements on the stability of points $\mathcal{H}_i$. We can conclude however that one of these points $\mathcal{H}_3$ is always a saddle.
As in the previous case, the stability of point $\mathcal{A}$ can be determined by the analysis of the central manifold. However, from the sign of the other eigenvalues, we can conclude that the point is unstable. In Figs.~\ref{Fig:stabR1}
and~\ref{Fig:stabR2} we also plot the stability, see  Table~\ref{TavolaRR3RBR}.

\newpage
\section{Sixth order terms vs. fourth order terms}\label{6thVS4th}
It is useful to compare the results that we have obtained so far with an analysis of fourth order models  made with the same approach (see also Ref.~\cite{Carloni:2015jla} for an equivalent, but slightly different choice of some of the dynamical variables). For simplicity we will consider here a fourth order theory of the form  $f= R+ \alpha R^{q}$. For this choice of $f$ the cosmological equations read
\begin{align}\label{CosmEq4}
\begin{split}
 H^{2}+\frac{k}{a^2} =&
\frac{1}{3(1+\alpha q R^{q-1})}\left\{\frac{1}{2}\alpha(q-1)R^q-3\alpha q (q-1) H R^{q-2}\dot{R}+\mu_{{m}}\right\}\,,\\
2\dot{H}+H^{2}+\frac{k}{a^2} =&
-\frac{1}{(1+\alpha q R^{q-1})}\left\{\frac{1}{2}\alpha(q-1)R^q-3\alpha q (q-1) H R^{q-2}\dot{R}\right.\\ &+q(q-1)(q-2)R^{q-3}\dot{R}^2+q(q-1)R^{q-2}R\ddot{R}+\,p_{{m}}\bigg\}\,,
\end{split}
\end{align}
Defining the variables
\begin{align} \label{DynVar4}
\begin{split}
\mathbb{R}=\frac{R}{6 H^2},\quad \mathbb{K}=\frac{k}{a^2 H^2},\quad \Omega =\frac{\mu }{3H^2},\\ \mathbb{J}=\mathfrak j,\quad \mathbb{Q}={\mathfrak q},\quad \mathbb{A}=\frac{R_0}{H^2}\,.
\end{split}
\end{align}
which are a subset of the variable in Eq.~\eqref{DynVar}, the cosmological equations can be written as
\begin{align}\label{DynSysRR^n4}
\begin{split}
 &\DerN{\mathbb{R}}=\frac{\mathbb{R}
   \left[q\left(2 n-3 \right)\K-\left(2 q^2 +3 q+1\right)\mathbb{R} +q \Omega+4 q^2 -5 q\right]}{q(q-1) }\\
   &~~~~~~~~-\frac{(K-\Omega +1)}{6^{q-1} \alpha  (q-1) q \A^{q-1}  \mathbb{R}^{q-2}},\\
 &\DerN{\Omega}=\Omega\left(1-3 w+2 \K-2\mathbb{R}\right),\\
 &\DerN{\K}=2 \K(\K- \mathbb{R}+1),\\ 
   &\DerN{\mathbb{A} }=2\mathbb{A} (2+\K-\mathbb{R}).
\end{split}
\end{align}
with the constraints
\begin{align}
\begin{split}
&  \mathbb{R}=\mathbb{K}+\mathbb{Q}+2,\\
& 6\left[(1+\mathbb{K}-\mathbb{R})\left(1+\frac{6^{q-1}\mathbb{R}}{\A^{q-1}}\right)+\mathbb{R}- \Omega\right]+\\
&+\alpha  6^q \mathbb{A}^{1-q}
   \mathbb{R}^{q-2} \left[q (q-1) \left( \mathbb{J}+\K^2-2 \K-4\right)+\right.\\
&\left.-2  q (q-1) \K \mathbb{R}+(q^2-q+1) \mathbb{R}^2\right]=0.
\end{split}
\end{align}
The solutions associated to the fixed points can be obtained from the equation
\begin{align}\label{RAy3Ord}
\begin{split}
& \mathfrak{s}=\frac{1}{H}\frac{ d^3{H}}{d N^3}
\end{split}
\end{align}
where $\mathfrak{s}$ is defined in Eq.~\eqref{HubbleVarDSN} and its expression in the fixed point can be deduced by the second of Eqs.~\eqref{CosmEq4} as we have done for the higher order case. As in the previous sections the solution can be given in general  noting that the characteristic polynomial for this equation has one real root and a pair of complex roots. Hence, we can write an exact solution for $H(N)$:
\begin{equation} \label{SolHFixPointsGen4}
\begin{split}
H=\exp \left(-p\, N\right)+\exp \left(\frac{1}{2}p N\right)\left[H \cos\left(p\frac{\sqrt{3}}{2} N\right)+\bar{H}\sin\left(p\frac{\sqrt{3}}{2} N\right)\right],\\
\end{split}
\end{equation}
where $p=-\sqrt[3]{\mathfrak{s}^*}$, $H$ and $\bar{H}$ are integration constants. Naturally for $\mathfrak{s}^*=0$ we have the usual equation for the scale factor
 \begin{equation}
\dot{a}=a\sum_{i=0}^{2}H_i (\ln a)^i.
\end{equation}
The fixed points for the system in Eq.~\eqref{DynSysRR^n4} with their stability is presented in Table~\ref{TavolaR+Rn}.

\begin{table}[h]
\begin{center}
\caption{Fixed points of the fourth order model $f(R)=R+\alpha R^{q}$ with their interval of existence and their associated solutions. Here A stands for attractor, R stands for repeller, S stands for saddle, and NHS for non-hyperbolic saddle.} \label{TavolaR+Rn}
\begin{tabular}{llllll} \hline\hline
Point & Coordinates $\{\mathbb{R},\mathbb{J},\mathbb{K},\Omega, \A\}$  & Solution & Existence & Stability
\\ \hline\\
$\mathcal{A}$ & $\left\{ 0, 1, -1 , 0,0\right\}$ & $\mathfrak s=-1$ & $\alpha\neq 0$ & S \\ \\
$\mathcal{B}$ & $\left\{ 0,4, 0, 0,0\right\}$ & $\mathfrak s=-8$& $\alpha\neq 0$& R or S \\ \\
$\mathcal{C}$ & $\left\{2, 0, 0, 0, 12\left[\alpha (q-2)\right]^{\frac{1}{q-1}}\right\}$   & $\mathfrak s=0$ &\makecell[l]{if $q\in\Re$\\ $\alpha(q-2)>0$} & A if $\frac{32}{25}\lesssim q<2$\\ \\
$\mathcal{D}$ & $\left\{ 2n(n-1), 1 , 2 (n-1) n-1, 0,0\right\}$  & $\mathfrak s=-1$& $q>1$ & S   \\ \\
$\mathcal{E}$ & $\left\{ \frac{(5-4 n) n}{4 n^2-6 n+2}, \left(\frac{n-2}{(n-1)(2n-1)}\right)^3, 0, 0,0\right\}$  & $\mathfrak s=\left(\frac{n-2}{(n-1)(2n-1)}\right)^3$&  $q>1$  & A if $q>2$  \\ \\
\hline\hline\\
 \end{tabular}
   \end{center}
\end{table}

Let us now repeat the same analysis for a theory that contains the fourth order term considered above {\it plus} a sixth order term. Consider then the action
\begin{equation}
{\mathcal A}=\int d^4x\sqrt{-g} \left[ R+ \alpha  R_0^{1-q} R^{q}+ \gamma R_0^{-2} R \Box R+{\cal L}_m\right]\,,
\end{equation}
which implies  $f_1=R_0 R+\alpha  R_0^{1-q} R^{q}$ and $f_2= R_0^{-2} R$. The non zero auxiliary quantities in Eq.~\rf{XYZT1} are
\begin{align}\label{XYZT1q}
\begin{split}
& {\bf X}_1\left(\mathbb{A},\mathbb{R}\right)= 6\mathbb{R}+\alpha 6^q\mathbb{R}^q\mathbb{A}^{1-q},\qquad {\bf X}_2\left(\mathbb{A},\mathbb{R} \right)=\frac{6\gamma \mathbb{R}}{\mathbb{A}^2},\\ 
&{\bf Y}_1\left(\mathbb{A},\mathbb{R} \right)= 1+6^{q-1}\alpha \mathbb{R}^{q-1}\mathbb{A}^{1-q}, \qquad {\bf Y}_2\left(\mathbb{A},\mathbb{R} \right)=\frac{\gamma}{\mathbb{A}^2}, \\
&{\bf Z}_1\left(\mathbb{A},\mathbb{R} \right) =\alpha q(q-1)6^{q-2} \mathbb{R}^{q-2}\mathbb{A}^{1-q},\qquad {\bf W}_1\left(\mathbb{A},\mathbb{R} \right) =\alpha q(q-1)(q-2)6^{q-3} \mathbb{R}^{q-3}\mathbb{A}^{1-q}.
\end{split}
\end{align}
As before the cosmological equations can be decoupled to give an explicit equation for
$\mathbb{S}_1$ and $\mathbb{S}_2$ and one can construct the dynamical system equations to have:
\begin{align}\label{DynSysRRqRBR}
\begin{split}
\DerN{\mathbb{R}}&=\mathbb{J}+(\mathbb{K}-2) \mathbb{K}-(\mathbb{R}-2)^2 ,\\ 
\DerN{\mathbb{B}}&=\mathbb{B} (3 \mathbb{K}-3 \mathbb{R}+7)+\frac{1}{2} \left(\mathbb{J}+\mathbb{K}^2-2 \mathbb{K}
   (\mathbb{R}+1)+\mathbb{R}^2-4\right)^2\\
&~~~-\frac{\alpha }{\gamma } \left\{2^{q-3} 3^{q-2} \A^{3-q} \mathbb{R}^{q-2} \left[q(q-1)  (\mathbb{J}+(\mathbb{K}-2)
   \mathbb{K}-4)+q \mathbb{R} (\mathbb{K} (3-2 q)+1)+(q-1)^2 \mathbb{R}^2\right]\right\}\\
  &~~~+\frac{\A^2}{12 \gamma } (-\mathbb{K}+\Omega -1),\\
\DerN{\Omega} &=\Omega  (1-3w+2 \mathbb{K}-2 \mathbb{R}),\\ 
 \DerN{\mathbb J} &=-\mathbb{B}+\mathbb{J} (5
   \mathbb{K}-5 \mathbb{R}+3)+(\mathbb{K}-\mathbb{R}) \left(\mathbb{K}^2-\mathbb{K} (2 \mathbb{R}+7)+\mathbb{R}
   (\mathbb{R}+5)\right)-22 \mathbb{K}+20 \mathbb{R}-12,\\
\DerN{\mathbb{K}}&=2 \mathbb{K} (\mathbb{K}-\mathbb{R}+1),\\
\DerN{\mathbb{A}}&=2 \mathbb{A}(\mathbb{K}-\mathbb{R}+2)\,.
\end{split}
\end{align}
In Table~\ref{TavolaRRqRBR} we give the fixed points and their stability.

\begin{table}[h]
\begin{center}
\caption{Fixed points of the model $f(R,\Box R)=R_0^{-1} R+ R_0^{1-q} R^q+ R_0^{-3} R \Box R$ and their  associated solutions. Here A stays for attractor, R for repeller, NHS for non hyperbolic saddle. The quantities $\mathbb{R}^*_i$ are the solutions of Eq.~\rf{EqPtH}. We assume $\{\alpha,\gamma\}\neq0$ and $q\neq 1$. } \label{TavolaRRqRBR}
\begin{tabular}{llllll} \hline\hline
Point & Coordinates  & Solution & Existence/ &  Stability \\
&$\{\mathbb{R},\mathbb{B},\mathbb{J},\Omega,\mathbb{K}, \A\}$ & parameter $\mathfrak{s}_2$  & Phsyical \\ \hline\\
$\mathcal{A}$ & $\left\{ 0,0,1,0, -1 , 0\right\}$  & $\mathfrak{s}_2=-1$ & $q\leq 3$ & NHS\\ \\
$\mathcal{B}$ & $\left\{ 0, 0, 4, 0, 0, 0\right\}$  & $\mathfrak{s}_2=-32$ & always & \makecell[l]{ R for $w < 1/3$ \\ S for $w > 1/3$}\\ \\
$\mathcal{C}$ & $\left\{ 2, 0, 0,0, 0, 12\left[\alpha (q-2)\right]^{\frac{1}{q-1}}\right\}$ & $\mathfrak{s}_2=0$ &\makecell[l]{if $q\in\Re$\\ $\alpha(q-2)>0$}& S\\ \\
$\mathcal{G}$ & $\left\{ 12+ \frac{2\gamma}{3\alpha}, 0, 1,0, 11+ \frac{2\gamma}{3\alpha}, 0\right\}$  & $\mathfrak{s}_2=\mathfrak{s}_\mathcal{G}$ & $q=3$ & S  \\ \\
$\mathcal{H}_1$ & $\left\{\mathbb{R}^*_1 , -6\mathbb{R}^*_1(\mathbb{R}^*_1-1)(\mathbb{R}^*_1-2),(\mathbb{R}^*_1-2)^2,0,0,0\right\}$  & $\mathfrak{s}_2=\sigma_{1}$  & q=3, Fig.~\ref{Fig:existPointH} &   Fig.~\ref{Fig:stabR1}  \\ \\
$\mathcal{H}_2$ & $\left\{\mathbb{R}^*_2 , -6\mathbb{R}^*_2(\mathbb{R}^*_2-1)(\mathbb{R}^*_2-2),(\mathbb{R}^*_2-2)^2,0,0,0\right\}$  & $\mathfrak{s}_2=\sigma_{2}$  & q=3, Fig.~\ref{Fig:existPointH} &   Fig.~\ref{Fig:stabR2}  \\ \\
$\mathcal{H}_3$ & $\left\{\mathbb{R}^*_3 , -6\mathbb{R}^*_3(\mathbb{R}^*_3-1)(\mathbb{R}^*_3-2),(\mathbb{R}^*_3-2)^2,0,0,0\right\}$  & $\mathfrak{s}_2=\sigma_{3}$  & q=3, Fig.~\ref{Fig:existPointH} &   S  \\ \\
$\mathcal{I}_{\pm}$ & $\left\{\frac{1}{10}\left(16\pm\sqrt{46}\right) , \frac{1}{50}\left(31\pm4\sqrt{46}\right),-\frac{9}{250}\left(74\pm 9\sqrt{46}\right),0,0,0\right\}$  & $\mathfrak{s}_2=\sigma_4$  & q<3 &   S  \\ \\
\hline \\ 
\multicolumn{5}{l}{$\mathfrak{s}_\mathcal{G}=-577-5184\frac{\alpha}{\gamma}-11\frac{\gamma}{\alpha}$} \\ \\
\multicolumn{5}{l}{\makecell[l]{$\sigma_i= \frac{\alpha}{\gamma}\left(-150 \mathbb{R}_{*,i}^4+435 \mathbb{R}_{*,i}^3-252 \mathbb{R}_{*,i}^2\right)+101
   \mathbb{R}_{*,i}^5-610 \mathbb{R}_{*,i}^4+1306 \mathbb{R}_{*,i}^3-1180 \mathbb{R}_{*,i}^2+416 \mathbb{R}_{*,i}-32\neq 0$ }}\\ \\
\multicolumn{5}{l}{ $\sigma_4=5^{-\left(q+1\right)}\left(\pm \frac{4549}{2}-\frac{2689 \sqrt{\frac{23}{2}}}{4}\right)$}\\ \\
\hline\hline\\
 \end{tabular}
   \end{center}
\end{table}

Although fundamentally different the two phase spaces present some similarities. Points~$\mathcal{A}$, $\mathcal{B}$ and $\mathcal{C}$ have exactly the same coordinates. In Points~$\mathcal{E}$  and Point~$\mathcal{D}$, instead, the relation among the values of some of the coordinates is  the same as the one of Points~$\mathcal{H}$. The difference in the coordinates of these points is probably due to the additional contributions generated in the gravitational field equations by the $R\Box R$ correction. As one could expect, the same additional terms can change the stability of all the fixed points.

\begin{figure}[h] 
\includegraphics[scale=0.6]{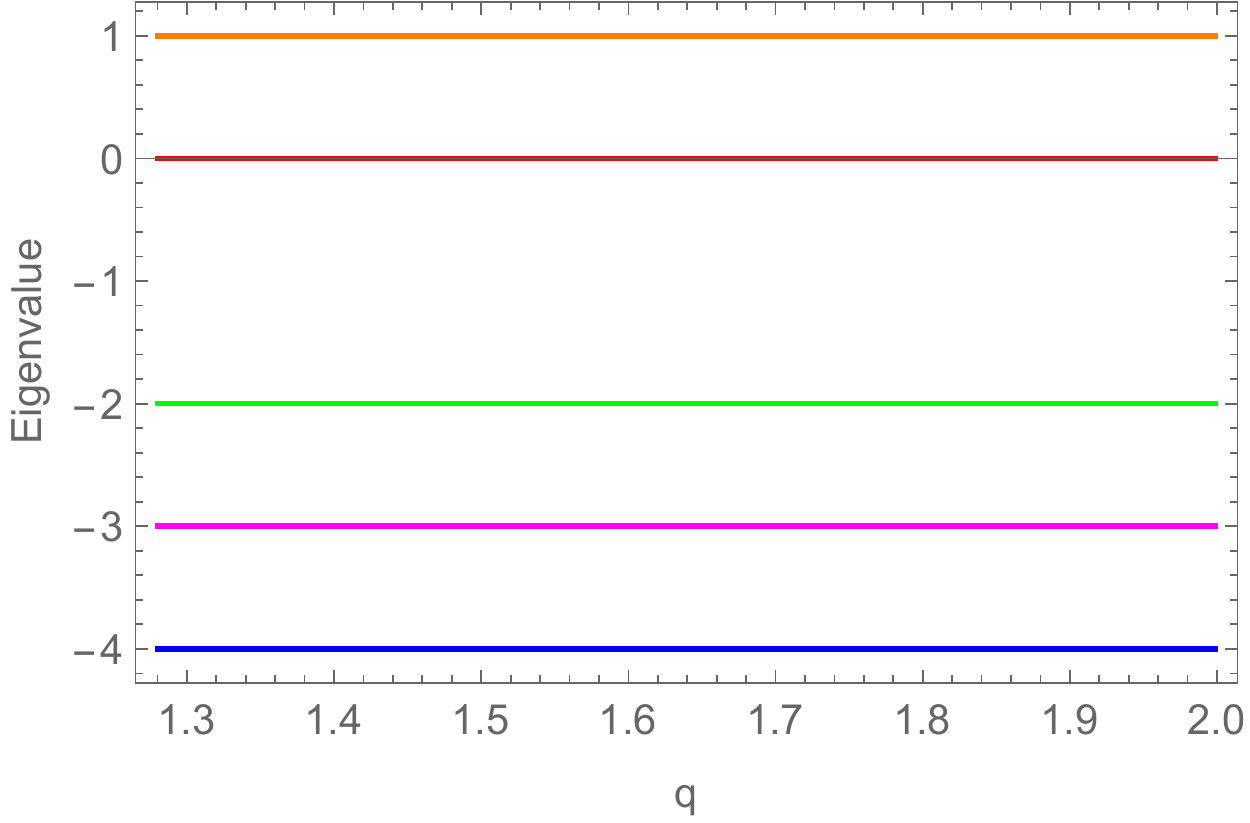}
\caption{Sign of the real part of the four eigenvalues associated to point $\mathcal{C}$ in the case $\alpha=1$, $\gamma=1$, $w=0$. The signs of the eigenvalues are discordant whatever the choice of the values of these parameters.}
\label{Fig:PointC6Ord}
\end{figure}

For our purposes, the most important result of this comparative analysis is the fact that both the phase spaces present the fixed point $\mathcal{C}$. As we have seen, such point is characterised  by the vanishing of the quantity associated to  both $\mathfrak{s}=0$ and $\mathfrak{s}_2=0$, and it can represent a solution with a finite time singularity. Looking at Table~\ref{TavolaR+Rn} we see that the fourth order theory  point $\mathcal{C}$ for $32/25<q<2$ is an attractor.  However, in the sixth order theory, it is possible to prove numerically that in the interval $32/25<q<2$ the point $\mathcal{C}$ is always unstable, see Fig.~\ref{Fig:PointC6Ord}. Therefore we can say that the introduction of the sixth order terms prevents the cosmology to evolve towards $\mathcal{C}$. Effectively, this amounts to ``curing'' the pathology of the fourth order model as the sixth order terms prevents the occurrence of a finite time singularity. In this sense, we can say that, as the time asymptotic state of sixth order cosmologies is never singular, these models are more ``stable'' with respect to the appearance of singularities. When we will consider eight order corrections, we will use in  the  results obtained in this section to reach the same conclusion.

\section{Going beyond sixth-order}
\subsection{The basic equations}\label{BasicEqbbb}

Let us start extending the set of variables used in the previous section, i.e., 
\begin{align} \label{DynVar2}
\begin{split}
&\mathbb{R}=\frac{R}{6 H^2},\quad \mathbb{B}=\frac{\Box R}{6H^4},\quad  \mathbb{K}=\frac{k}{a^2 H^2},\quad \Omega =\frac{\mu }{3H^2},\quad \mathbb{J}={\mathfrak j},\\ &\mathbb{Q}={\mathfrak q},\quad \mathbb{S}={\mathfrak s},\quad \mathbb{S}_1={\mathfrak s}_1 ,\quad \mathbb{S}_2={\mathfrak s}_2, \quad \mathbb{S}_3={\mathfrak s}_3
,\quad \mathbb{A}=\frac{R_0}{H^2}\,.
\end{split}
\end{align}
The Jacobian of this variable definition reads
\begin{align} \label{Jac8}
M_8=-\frac{1}{108 a^2 H^{47}},
\end{align}
which implies that, as in the sixth-order case, the variables are always regular if
$H\neq 0$ and $a \neq 0$.

The requirement to have a closed systems of equations implies the introduction of the auxiliary quantities,
\begin{align}\label{ParHO1}
\begin{split}
& {\bf X}\left(\mathbb{A},\mathbb{R},\mathbb{B}\right)= \frac{f\left(\mathbb{A},\mathbb{R},\mathbb{B}\right)}{H^2},\\ 
&{\bf Y}_1\left(\mathbb{A},\mathbb{R},\mathbb{B} \right)= f^{(1,0)}_{R,\Box R}\left(\mathbb{A},\mathbb{R},\mathbb{B}\right),\\ 
&{\bf Y}_2\left(\mathbb{A},\mathbb{R},\mathbb{B} \right)= H^4 f^{(1,1)}_{R,\Box R}\left(\mathbb{A},\mathbb{R},\mathbb{B}\right),\\ 
&{\bf Y}_3\left(\mathbb{A},\mathbb{R},\mathbb{B} \right)= H^2 f^{(0,1)}_{R,\Box R}\left(\mathbb{A},\mathbb{R},\mathbb{B}\right),\\ 
&{\bf Z}_1\left(\mathbb{A},\mathbb{R},\mathbb{B}\right)= H^2 f^{(2,0)}_{R,\Box R}\left(\mathbb{A},\mathbb{R},\mathbb{B}\right),\\
&{\bf Z}_2\left(\mathbb{A},\mathbb{R},\mathbb{B} \right)=H^6 f^{(2,1)}_{R,\Box R}\left(\mathbb{A},\mathbb{R},\mathbb{B}\right),\\
& {\bf Z}_3\left(\mathbb{A},\mathbb{R},\mathbb{B}\right)= H^{10} f^{(2,2)}_{R,\Box R}\left(\mathbb{A},\mathbb{R},\mathbb{B}\right),\\ 
& {\bf Z}_4\left(\mathbb{A},\mathbb{R},\mathbb{B} \right)= H^8 f^{(1,2)}_{R,\Box R}\left(\mathbb{A},\mathbb{R},\mathbb{B}\right),\\
&{\bf Z}_5\left(\mathbb{A},\mathbb{R},\mathbb{B} \right)= H^6 f^{(0,2)}_{R,\Box R}\left(\mathbb{A},\mathbb{R},\mathbb{B}\right),
\end{split}
\end{align}
\begin{align}\label{ParHO2}
\begin{split}
& {\bf W}_1\left(\mathbb{A},\mathbb{R},\mathbb{B}\right)=H^4 f^{(3,0)}_{R,\Box R}\left(\mathbb{A},\mathbb{R},\mathbb{B}\right),\\ 
&{\bf W}_2\left(\mathbb{A},\mathbb{R},\mathbb{B} \right)= H^8 f^{(3,1)}_{R,\Box R}\left(\mathbb{A},\mathbb{R},\mathbb{B}\right),\\ 
&{\bf W}_3\left(\mathbb{A},\mathbb{R},\mathbb{B}\right)= H^{12} f^{(3,2)}_{R,\Box R}\left(\mathbb{A},\mathbb{R},\mathbb{B}\right), \\ 
& {\bf W}_4\left(\mathbb{A},\mathbb{R} ,\mathbb{B}\right)= H^{14} f^{(2,3)}_{R,\Box R}\left(\mathbb{A},\mathbb{R},\mathbb{B}\right),\\ 
&{\bf W}_5\left(\mathbb{A},\mathbb{R},\mathbb{B}\right)= H^{12} f^{(1,3)}_{R,\Box R}\left(\mathbb{A},\mathbb{R},\mathbb{B}\right),\\
&{\bf W}_6\left(\mathbb{A},\mathbb{R},\mathbb{B} \right)= H^{10} f^{(0,3)}_{R,\Box R}\left(\mathbb{A},\mathbb{R},\mathbb{B}\right),\\
&{\bf T}_1\left(\mathbb{A},\mathbb{R},\mathbb{B}\right)= H^{14} f^{(0,4)}_{R,\Box R}\left(\mathbb{A},\mathbb{R},\mathbb{B}\right),\\
&{\bf T}_2\left(\mathbb{A},\mathbb{R},\mathbb{B} \right)= H^{16} f^{(1,4)}_{R,\Box R}\left(\mathbb{A},\mathbb{R},\mathbb{B}\right),\\
&{\bf T}_3\left(\mathbb{A},\mathbb{R},\mathbb{B}\right)= H^{10} f^{(4,1)}_{R,\Box R}\left(\mathbb{A},\mathbb{R},\mathbb{B}\right),\\ 
&{\bf V}\left(\mathbb{A},\mathbb{R},\mathbb{B}\right)= H^{18} f^{(0,5)}_{R,\Box R}\left(\mathbb{A},\mathbb{R},\mathbb{B}\right),
\end{split}
\end{align}
where, for simplicity, we indicate with $f^{(i,j)}_{R,\Box R}$ the $i$-th  $R$-derivative and the $j$-th $\Box R$-derivative of $f$.

The cosmological dynamics can be  described by the autonomous system 
\begin{align}\label{DynSys8}
\begin{split}
&\DerN{\mathbb{R}}=\mathbb{J}-2 \mathbb{K}-2 \mathbb{Q} \mathbb{R}+\mathbb{Q} (\mathbb{Q}+4) ,\\ 
&\DerN{\mathbb{B}}= -4 \mathbb{B} \mathbb{Q}-4 \mathbb{J}^2+\mathbb{J} (2 \mathbb{K}-\mathbb{Q} (11
   \mathbb{Q}+43)-12)\\ 
&~~~~~~~~~~-\mathbb{Q} (\mathbb{Q} (-2 \mathbb{K}+\mathbb{Q} (\mathbb{Q}+22)+36)+7 \mathbb{S})\\ 
&~~~~~~~~~~-4\mathbb{K}-7 \mathbb{S}-\mathbb{S}_1,\\ 
&\DerN{\Omega} =- \Omega  (2 \mathbb{Q}+3w+3),\\ 
&\DerN{\mathbb J}=\mathbb{S}-\mathbb{J} \mathbb{Q},\qquad
\DerN{\mathbb{Q}}=\mathbb{J}-\mathbb{Q}^2,\\ 
& \DerN{\mathbb{K}}=-2 \mathbb{K} (\mathbb{Q}+1),\qquad \DerN{\mathbb{S}}=\mathbb{S}_1-\mathbb{Q} \mathbb{S},\\ 
&\DerN{\mathbb{S}_1}=\mathbb{S}_2-\mathbb{Q}  \mathbb{S}_1,\qquad 
\DerN{\mathbb{S}_2}=\mathbb{S}_3({\bf X},{\bf Y}_1,..)-\mathbb{Q}  \mathbb{S}_2,\\ 
&\DerN{\mathbb{S}_3}=\mathbb{S}_4({\bf X},{\bf Y}_1,..)-\mathbb{Q}  \mathbb{S}_3({\bf X},{\bf Y}_1,..), 
\\& \DerN{\mathbb{A}}=-2 A\mathbb{Q}\,,
\end{split}
\end{align}
where  $\mathbb{S}_4={\mathfrak s}_4$. As before, the system above is completed by three constraints: the one coming from the modified Friedmann equation, Eq.~\rf{CosmicEq}, and the ones in Eq.~\rf{ConstrRBR}. We choose to use these constraints to eliminate $\mathbb{Q}, \mathbb{S}$, and $\mathbb{S}_3$. The variable $\mathbb{S}_4$ instead, can substituted using the modified Raychaudhuri equation. In
Eq.~\eqref{DynSys8} these variables are not substituted explicitly in order to give a more compact representation of the system. The substitution of $\mathbb{S}_3$ and $\mathbb{S}_4$ also brings in the system the parameters  given in Eqs.~\eqref{ParHO1} and \eqref{ParHO2}.

In the same way of Sec.~\ref{BasicEqb}, the solutions associated to the fixed points can be found by solving the differential equation
\begin{align}\label{RAyHOrd7}
\begin{split}
& \frac{1}{H}\frac{ d^7{H}}{d N^7}=\mathfrak{s}^*_4,
\end{split}
\end{align}
where $\mathfrak{s}^*_4$ is provided by the modified Raychaudhuri equation, Eq.~\rf{CosmicEq2}.

Equation~\rf{RAyHOrd7} can be shown to give a result structurally similar to the one of the previous section.  The characteristic polynomial of Eq.~\rf{RAy6Ord} has one real and three pairs of complex roots. This leads to the exact solution
\begin{equation} \label{SolHFixPointsGen8}
\begin{split}
H=\sum_{i=0}^{3}\exp \left(p\,\alpha_i N\right)\left[H_i \cos\left(\beta_i p N\right)+\bar{H}_i \sin\left(\beta_i p N\right)\right],\\
\end{split}
\end{equation}
where $p=-\sqrt[7]{\mathfrak{s}^*_4}$, $H_i$ and $\bar{H}_i$ are integration constants and $a_i$ and $b_i$ are the real and imaginary part of the seventh root of the unity. These quantities are expressed by the relation
\begin{equation}
\begin{array}{ll}
\alpha_0=-1, & \beta_0=0,\\
\alpha_i=\frac{r}{2}, &i\neq0,\\
\beta_i=\sqrt{1-\frac{r^2}{4}}, &i\neq0,\,
\end{array}
\end{equation}
where $r$ is the solution of the algebraic equation $r^3+r^2-2r-1=0$. The scale factor is given by the equation
\begin{equation} \label{EqaFixPoints8}
\dot{a}=\sum_{i=0}^{3}a^{1+p\,\alpha_i}\left[H_i \cos\left(\beta_i p \ln a\right)+\bar{H}_i \sin\left(\beta_i p \ln a\right)\right],\\
\end{equation} 
which can be solved numerically. As before, $H$ and $a$ are parameterised only by the quantity $p$ i.e. $\mathfrak{s}^*_4$. In the following we will characterise these solutions only by the value of 
$\mathfrak{s}^*_4$. 

\begin{figure}[h] 
\includegraphics[scale=0.6]{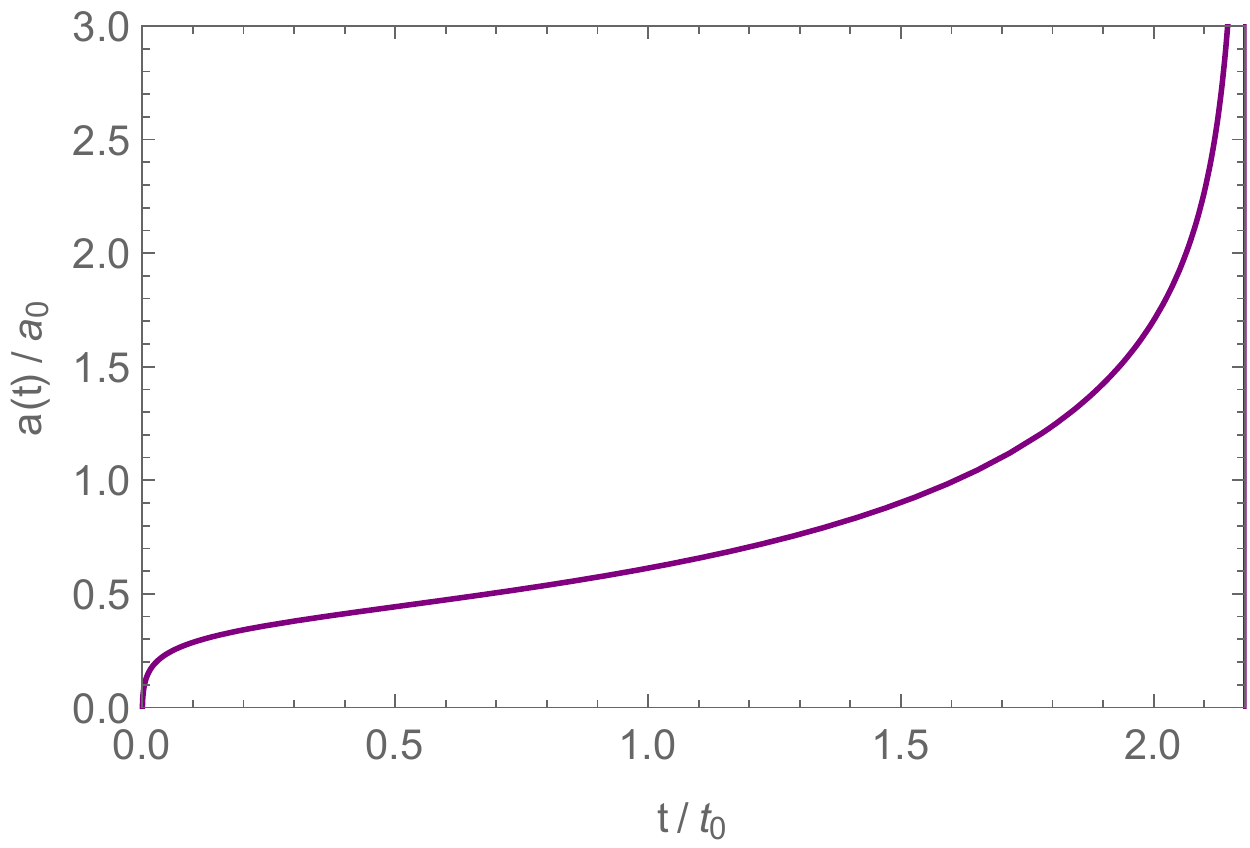}
\caption{Behaviour of the scale factor in a fixed point with $\mathfrak{s}^*_4=0$. All integration constants have been chosen to be one.}
\label{Fig:POINT_C_8TH_LIN}
\end{figure}

If $\mathfrak{s}^*_4=0$ then Eq.~\rf{RAyHOrd7} can be written as
\begin{equation}\label{SolHn}
\dot{N}=\sum_{i=0}^{6}H_i N^i,
\end{equation}
and the existence of a finite time singularity is only possible if the polynomial on the
left hand side has  complex roots. In Fig.~\ref{Fig:POINT_C_8TH_LIN} we show the time dependence of the scale factor corresponding to fixed points with $\mathfrak{s}^*_4=0$.

We will consider now three examples of theories of order eight. As in the previous section we will first examine a model in which the Hilbert-Einstein term appears together with a contribution of order eight. In the second we will introduce a fourth order terms in order to explore the interaction of the eight-order terms with the fourth-order ones. Finally in the third awe will explore a theory in which the Hilbert-Einstein appears together with fourth, sixth and eight orders.

\subsection{Three examples}\label{twoa}

\subsubsection{Case $f= R+ \gamma(\Box R)^2$}\label{Example8_1}
Here we examine a model in which the only
new  contribution comes from a term of order eight.
The action for this theory can be written as 
\begin{equation}\label{ActionRBRBR}
{\mathcal A}=\int d^4x\sqrt{-g} \left[R+ \gamma R_0^{-3}  (\Box R)^2+{\cal L}_m\right]\,.
\end{equation}
For this theory the only non zero auxiliary quantities in Eq.~\rf{XYZT1} are
\begin{align}\label{XYZT1w}
\begin{split}
& {\bf X}\left(\mathbb{A},\mathbb{R},\mathbb{B}\right)= 6\left(\mathbb{R}+\frac{6\gamma\mathbb{B}}{\mathbb{A}^2}\right), \qquad{\bf Y}_1\left(\mathbb{A},\mathbb{R},\mathbb{B} \right)= 1, \\
&{\bf Y}_3\left(\mathbb{A},\mathbb{R} ,\mathbb{B}\right)= \frac{12\gamma\mathbb{B}}{\mathbb{A}^3},\qquad {\bf Z}_5\left(\mathbb{A},\mathbb{R},\mathbb{B} \right) =\frac{2\gamma}{\mathbb{A}^3}.
\end{split}
\end{align}
The cosmological equations can be decoupled to give explicit equations for $\mathbb{S}_3$ and $\mathbb{S}_4$, which can be found
in Eq.~(\ref{c1c1}) of Appendix \ref{App}.

The  dynamical system Eq.~\rf{DynSys8} then becomes
\begin{align}
\begin{split}
\DerN{\mathbb{R}}&=-(\mathbb{R}-2)^2+\mathbb{J}+(\mathbb{K}-2) \mathbb{K},\qquad \DerN{\mathbb{A}}=2  \mathbb{A}(\mathbb{K}-\mathbb{R}+2),\\ 
\DerN{\Omega}&=(1-3 w-2 \mathbb{K}+2 \mathbb{R})
   \Omega,\qquad \DerN{\mathbb{K}}=2 \mathbb{K} (\mathbb{K}-\mathbb{R}+1), \\
   \DerN{\mathbb{J}}&=\mathbb{B}+\mathbb{J} (-5 \mathbb{K}+5 \mathbb{R}-3)+22 \mathbb{K}+12,\qquad    \DerN{\mathbb{S}_1}=\left(2 + \mathbb{K} - \mathbb{R}\right) \mathbb{S}_1 + \mathbb{S}_2,\\\DerN{\mathbb{B}}&=6 \mathbb{K}^4-24 \mathbb{R} \mathbb{K}^3-26
   \mathbb{K}^3+36 \mathbb{R}^2 \mathbb{K}^2+66 \mathbb{R} \mathbb{K}^2-123 \mathbb{K}^2-24 \mathbb{R}^3 \mathbb{K}-54 \mathbb{R}^2
   \mathbb{K}\\
&~~~+226 \mathbb{R} \mathbb{K}-146 \mathbb{K}+6 \mathbb{R}^4+14 \mathbb{R}^3 -4 \mathbb{J}^2-103 \mathbb{R}^2+136 \mathbb{R}+\mathbb{B}  (-3 \mathbb{K}+3 \mathbb{R}+1)\\
&~~~+\mathbb{J} \left(17 \mathbb{K}^2-34 \mathbb{R} \mathbb{K}+36 \mathbb{K}+17 \mathbb{R}^2-34
   \mathbb{R}+37\right)-\mathbb{S}_1-68,\\
   \DerN{\mathbb{S}_2}&=\mathbb{S}_2 f_1(\mathbb{R},\mathbb{B},\mathbb{J},\mathbb{S}_1,\mathbb{K}, \A) + \frac{\alpha}{\gamma} f_2(\mathbb{R},\mathbb{B},\mathbb{J},\mathbb{S}_1,\mathbb{K}, \A) \\
&~~~+ \frac{\beta}{\gamma} f_3(\mathbb{R},\mathbb{B},\mathbb{J},\mathbb{S}_1,\mathbb{K}, \A)+  \frac{1}{\gamma}f_3(\mathbb{R},\mathbb{B},\mathbb{J},\mathbb{S}_1,\mathbb{K}, \A)
\,,
\end{split}
\end{align}
where the full equation for $\mathbb{S}_2$ is only shown in its structure due to it length. Its full expression can be found in Appendix~\ref{App}.

\begin{table*}
\begin{center}
\caption{Fixed points of $f(R,\Box R)= R+ \gamma (\Box R)^2$ and their  associated solutions. Here R stands for repeller, S for saddle, F$_A$  for attractive focus, NHS  for non hyperbolic saddle.} \label{TavolaRBRBR}
\begin{tabular}{llllll} \hline\hline
Point & Coordinates $\{\mathbb{R},\mathbb{B},\mathbb{J},\mathbb{S}_1,\mathbb{S}_2,\mathbb{K},\Omega, \A\}$ & Solution & Existence&  Stability \\ \hline\\
$\mathcal{A}$ & $\left\{ 1,0,1,1, -1,0, 0 , 0\right\}$  & $\mathfrak{s}_4=-1$ & $\gamma\neq 0$ & NHS \\ \\
$\mathcal{B}$ & $\left\{0,0, 4, 16 , -32, 0, 0, 0\right\}$  & $\mathfrak{s}_4=-128$ & $\gamma\neq 0$ & \makecell[c]{$R$ for $w<1/3$\\$S$ for $w>1/3$}\\ \\
$\mathcal{C}$ & $\left\{2,0, 0,0,0, 0, 0, 0\right\}$  & $\mathfrak{s}_4=0$ & $\gamma\neq 0$ & NHS\\ \\
$\mathcal{I}_1$ & $\left\{ a_\mathcal{H}^-,  -6a_\mathcal{H}^-(a_\mathcal{H}^--1)(a_\mathcal{H}^--2),(a_\mathcal{H}^--2)^2,(a_\mathcal{H}^--2)^4,(a_\mathcal{H}^--2)^5,0,0,0\right\}$&$\mathfrak{s}_4\approx -7.8\times 10^{-3}$ & $\gamma\neq 0$ & S\\ 
&&&&\\\\
$\mathcal{I}_2$ & $\left\{a_\mathcal{H}^+,  -6a_\mathcal{H}^+(a_\mathcal{H}^+-1)(a_\mathcal{H}^+-2),(a_\mathcal{H}^+-2)^2,(a_\mathcal{H}^+-2)^4,(a_\mathcal{H}^+-2)^5,0,0,0\right\}$&$\mathfrak{s}_4\approx 5.6\times 10^{-5}$ & $\gamma\neq 0$ & F$_A$\\ 
&&&&\\\\

\hline 
Line & Coordinates $\{\mathbb{R},\mathbb{B},\mathbb{J},\mathbb{S}_1,\mathbb{S}_2,\mathbb{K},\Omega, \A\}$ & Solution & Existence&  Stability \\ \hline
\multirow{2}{*}{$\mathcal{L}$} & \multirow{2}{*}{$\left\{ \mathbb{R}_*, 0, 1, 1, -1,0, \mathbb{R}_*-1, 0\right\}$ } & \multirow{2}{*}{$\mathfrak{s}_4=-1$}&\multirow{2}{*}{always}&\multirow{2}{*}{NHS}  \\
  && && \\\hline\\
\multicolumn{5}{c}{\makecell[c]{$a_\mathcal{I}^\pm=\frac{1}{210}\left(373\pm\sqrt{9769}\right) $}}\\\\\hline\hline\\
 \end{tabular}
   \end{center}
\end{table*}

\begin{figure}[!ht]
  \includegraphics[width=0.5\textwidth]{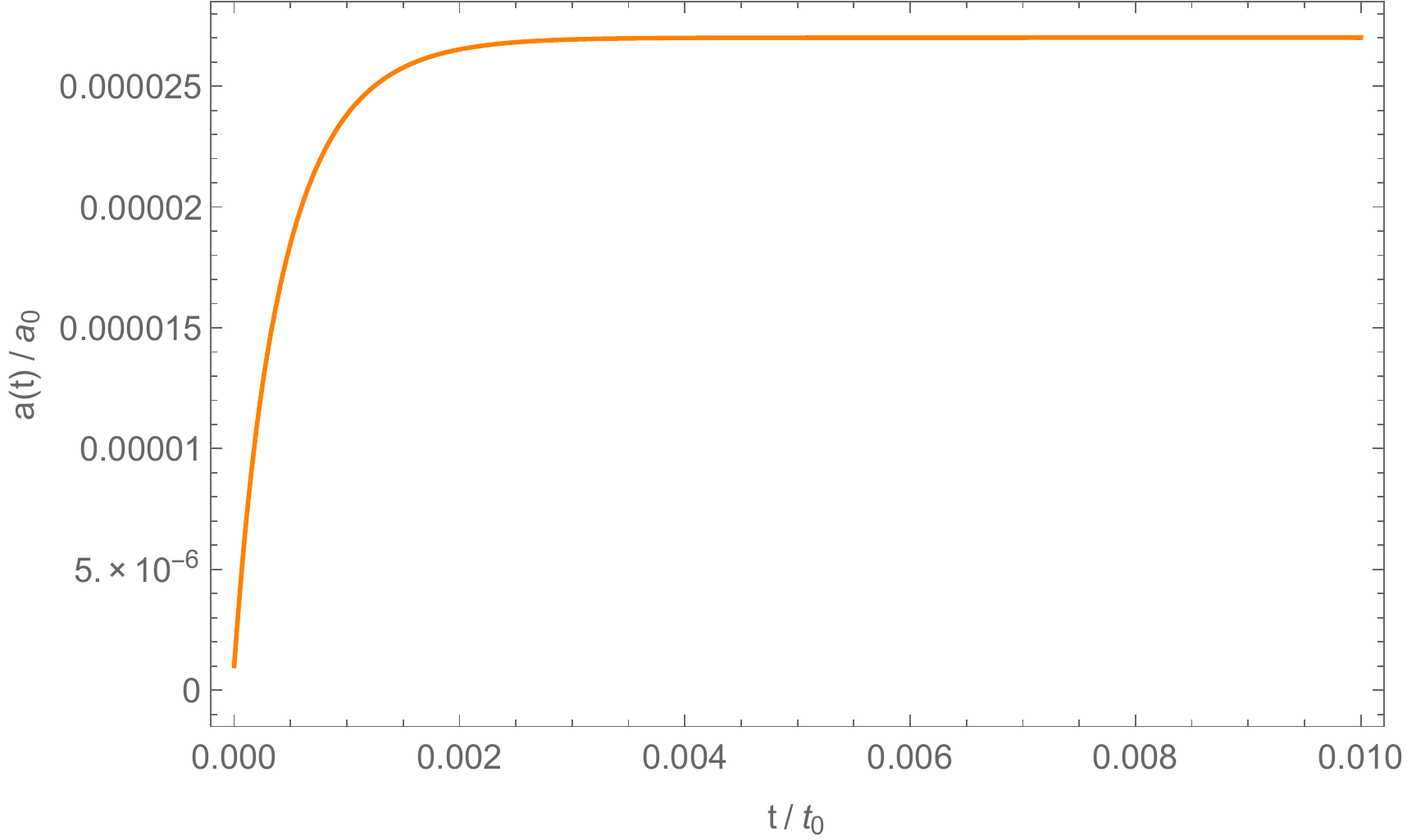}
   \caption{Behaviour of the scale factor in the fixed points of the phase space of the theory $f(R,\Box R)=R_0^{-1} R+ R_0^{-3} (\Box R)^2$. The integration constants have been chosen to be one.}
   \label{fig_ex8-1}
  \end{figure}

The system above presents the  invariant submanifolds ($\mathbb{A}=0$, $ \Omega=0$, $\mathbb{K}=0$) and therefore no global attractor with coordinates different from $\mathbb{A}=0$, $ \Omega=0$, $\mathbb{K}=0$ can exist. Table \ref{TavolaRBRBR} summarises the fixed points for this system with the associated solution and their stability.
 The system presents a line of fixed points, all unstable, and a global attractor, point $\mathcal{I}_2$, which is associated with a solution with non zero $\mathfrak{s}_4$. The solutions  for the scale factor are not structurally different form the ones of the sixth order case. In Fig.~\ref{fig_ex8-1} we give, as an example, a plot of the solution associated to $\mathcal{I}_2$. Points $\mathcal{A}$ and  $\mathcal{C}$ are non hyperbolic, the latter having two zero eigenvalues, but they can be both considered unstable. A detailed treatment of the stability of $\mathcal{C}$ would require blow up techniques.
We refer the reader to Ref.~\cite{DumortierBook} for more information on this topic.

\subsubsection{Case $f= R+ \alpha R^q +\gamma(\Box R)^2$}\label{Example8_2}

Here we consider the case
in which both eight-order and fourth-order terms are present.
The action for this theory can be written as 
\begin{equation}\label{ActionR3BRBR}
{\mathcal A}=\int d^4x\sqrt{-g} \left[R+\alpha R_0^{1-q} R^q + \gamma R_0^{-6}  (\Box R)^2+{\cal L}_m\right]\,.
\end{equation}
For this theory the only non zero auxiliary quantities in Eq.~\rf{XYZT1} are
\begin{align}\label{XYZT1z}
\begin{split}
& {\bf X}\left(\mathbb{A},\mathbb{R},\mathbb{B}\right)= 6\left(\mathbb{R}+\frac{6\gamma\mathbb{B}}{\mathbb{A}^2}+\alpha 6^q\mathbb{R}^q\mathbb{A}^{1-q}\right), \\
&{\bf Y}_1\left(\mathbb{A},\mathbb{R},\mathbb{B} \right)= 1+6^{q-1}\alpha \mathbb{R}^{q-1}\mathbb{A}^{1-q}, \\
&{\bf Y}_3\left(\mathbb{A},\mathbb{R} ,\mathbb{B}\right)= \frac{12\gamma\mathbb{B}}{\mathbb{A}^3},\\
& {\bf Z}_1\left(\mathbb{A},\mathbb{R},\mathbb{B} \right) =\alpha q(q-1)6^{q-2} \mathbb{R}^{q-2}\mathbb{A}^{1-q},\\
& {\bf Z}_5\left(\mathbb{A},\mathbb{R},\mathbb{B} \right) =\frac{2\gamma}{\mathbb{A}^3},\\
&{\bf W}_1\left(\mathbb{A},\mathbb{R} ,\mathbb{B}\right) =\alpha q(q-1)(q-2)6^{q-3} \mathbb{R}^{q-3}\mathbb{A}^{1-q}.
\end{split}
\end{align}
As before, the cosmological equations can be decoupled to give an explicit equation for $\mathbb{S}_3$ and  another for $\mathbb{S}_4$. However, we will not show them here due to their size.

The  dynamical system Eq.~\rf{DynSys8} is now
\begin{align}
\begin{split}
\DerN{\mathbb{R}}&=-(\mathbb{R}-2)^2+\mathbb{J}+(\mathbb{K}-2) \mathbb{K},\qquad \DerN{\mathbb{A}}=2  \mathbb{A}(\mathbb{K}-\mathbb{R}+2),\\ 
\DerN{\Omega}&=(1-3 w-2 \mathbb{K}+2 \mathbb{R})
   \Omega,\qquad \DerN{\mathbb{K}}=2 \mathbb{K} (\mathbb{K}-\mathbb{R}+1), \\
   \DerN{\mathbb{J}}&=\mathbb{B}+\mathbb{J} (-5 \mathbb{K}+5 \mathbb{R}-3)+22 \mathbb{K}+12,\qquad
   \DerN{\mathbb{S}_1}=\left(2 + \mathbb{K} - \mathbb{R}\right) \mathbb{S}_1 + \mathbb{S}_2,\\\DerN{\mathbb{B}}&=6 \mathbb{K}^4-24 \mathbb{R} \mathbb{K}^3-26
   \mathbb{K}^3+36 \mathbb{R}^2 \mathbb{K}^2+66 \mathbb{R} \mathbb{K}^2-123 \mathbb{K}^2-24 \mathbb{R}^3 \mathbb{K}-54 \mathbb{R}^2
   \mathbb{K}\\
&~~~+226 \mathbb{R} \mathbb{K}-146 \mathbb{K}+6 \mathbb{R}^4+14 \mathbb{R}^3 -4 \mathbb{J}^2-103 \mathbb{R}^2+136 \mathbb{R}+\mathbb{B}  (-3 \mathbb{K}+3 \mathbb{R}+1)\\
&~~~+\mathbb{J} \left(17 \mathbb{K}^2-34 \mathbb{R} \mathbb{K}+36 \mathbb{K}+17 \mathbb{R}^2-34
   \mathbb{R}+37\right)-\mathbb{S}_1-68,\\
\DerN{\mathbb{S}_2}&=\mathbb{S}_2 f_1(\mathbb{R},\mathbb{B},\mathbb{J},\mathbb{S}_1,\mathbb{K}, \A) + \frac{\alpha}{\gamma} f_2(\mathbb{R},\mathbb{B},\mathbb{J},\mathbb{S}_1,\mathbb{K}, \A) \\
&~~~+ \frac{\beta}{\gamma} f_3(\mathbb{R},\mathbb{B},\mathbb{J},\mathbb{S}_1,\mathbb{K}, \A)+  \frac{1}{\gamma}f_3(\mathbb{R},\mathbb{B},\mathbb{J},\mathbb{S}_1,\mathbb{K}, \A)
\,,
\end{split}
\end{align}
where the full equation for $\mathbb{S}_2$ is only shown in its structure due to its length. Its full expression can be found in Appendix~\ref{App}.

This system presents analogies with the ones of the previous examples. The invariant submanifolds  present  in these cases are also $\mathbb{A}=0$, $ \Omega=0$, $\mathbb{K}=0$ and therefore the only possible type global attractor must lay on the intersection of these coordinates. The fixed point with their stability and the parameter $\mathfrak{s}_4$ that characterise the solution is given in Table \ref{TavolaRRqBRBR}. The coordinates of the point $\mathcal{H}_i$ are determined by the solution of the equation
\begin{equation}\label{8Ord_q_conditon}
\frac{3 \alpha }{\gamma } (21 \mathbb{R}-44) \mathbb{R}^3+\mathbb{R}(\mathbb{R}-2)^2
   (\mathbb{R}-1) \left(105 \mathbb{R}^2-373 \mathbb{R}+308\right) =0\,.
\end{equation}
One of these points is an attractor for specific values of $\alpha$ and $\gamma$, see Table \ref{TavolaRRqBRBR}. In the other cases no attractor can be found in the finite phase space.
  
\begin{table*}
\begin{center}
\caption{Fixed points of $f(R,\Box R)= R+ \alpha R^q +\gamma (\Box R)^2$ and their  associated solutions.
Here A stands for Attractor, Re stands for Repeller, S stands for saddle, and NHS for non-hyperbolic saddle. We also assume $\alpha,\gamma \neq 0$. The index ``$i$'' of the points $\mathcal{H}_i$ runs from 1 to 5. The value of $\mathbb{R}^*_i$ are the roots of the Eq.~\eqref{8Ord_q_conditon}.}
\label{TavolaRRqBRBR}
\begin{tabular}{llllll} \hline\hline
Point & Coordinates $\{\mathbb{R},\mathbb{B},\mathbb{J},\mathbb{S}_1,\mathbb{S}_2,\mathbb{K},\Omega, \A\}$ & Solution & Existence&  Stability \\ \hline\\
$\mathcal{A}$ & $\left\{ 1,0,1,1, -1,0, 0 , 0\right\}$  & $\mathfrak{s}_4=-1$ & $q\leq3$ & NHS \\ \\
$\mathcal{B}$ & $\left\{0,0, 4, 16 , -32, 0, 0, 0\right\}$  & $\mathfrak{s}_4=-128$ & $\gamma\neq 0$ & \makecell[c]{$R$ for $w<1/3$\\$S$ for $w>1/3$}\\ \\
$\mathcal{C}$ & $\left\{2,0, 0,0,0, 0, 0,12\left[\alpha (q-2)\right]^{\frac{1}{q-1}} \right\}$  & $\mathfrak{s}_4=0$ &\makecell[l]{if $q\in\Re$\\ $\alpha(q-2)>0$}& S\\ \\
$\mathcal{I}_1$ & $\left\{2,0, 0,0,0, 0, 0, 0\right\}$  & $\mathfrak{s}_4=0$ & $\gamma\neq 0$ & NHS\\ \\
$\mathcal{H}_i$ & $\left\{\mathbb{R}^*_i , -6\mathbb{R}^*_i(\mathbb{R}^*_i-1)(\mathbb{R}^*_i-2),(\mathbb{R}^*_i-2)^2,(\mathbb{R}^*_i-2)^4,(\mathbb{R}^*_i-2)^5,0,0,0\right\}$  & $\mathfrak{s}_2=\sigma_{i}$  & q=4& \makecell[l]{One A for\\$\left|\frac{\alpha}{\gamma}\right|\gtrsim 0.011$\\$\left|\frac{\alpha}{\gamma}\right|\lesssim 0.0035$\\ other points\\ unstable}\\ \\
$\mathcal{I}_1$ & $\left\{ a_\mathcal{H}^-,  -6a_\mathcal{H}^-(a_\mathcal{H}^--1)(a_\mathcal{H}^--2),(a_\mathcal{H}^--2)^2,(a_\mathcal{H}^--2)^4,(a_\mathcal{H}^--2)^5,0,0,0\right\}$&$\mathfrak{s}_4\approx -7.8\times 10^{-3}$ & $q\leq3$ & S\\ 
&&&&\\\\
$\mathcal{I}_2$ & $\left\{a_\mathcal{H}^+,  -6a_\mathcal{H}^+(a_\mathcal{H}^+-1)(a_\mathcal{H}^+-2),(a_\mathcal{H}^+-2)^2,(a_\mathcal{H}^+-2)^4,(a_\mathcal{H}^+-2)^5,0,0,0\right\}$&$\mathfrak{s}_4\approx 5.6\times 10^{-5}$ & $q\leq3$& S\\ 
&&&&\\\\
\hline 
Line & Coordinates $\{\mathbb{R},\mathbb{B},\mathbb{J},\mathbb{S}_1,\mathbb{S}_2,\mathbb{K},\Omega, \A\}$ & Solution & Existence&  Stability \\ \hline
\multirow{2}{*}{$\mathcal{L}$} & \multirow{2}{*}{$\left\{ \mathbb{R}_*, 0, 1, 1, -1,0, \mathbb{R}_*-1, 0\right\}$ } & \multirow{2}{*}{$\mathfrak{s}_4=-1$}&\multirow{2}{*}{$q\leq3$}&\multirow{2}{*}{S}  \\
  && && \\\hline\\
\multicolumn{5}{c}{\makecell[c]{$a_\mathcal{H}^\pm=\frac{1}{210}\left(373\pm\sqrt{9769}\right) $}}\\\\\hline\hline\\
 \end{tabular}
   \end{center}
\end{table*}

\subsubsection{Case $f= R+ \alpha R^4 +\beta R \Box R + \gamma(\Box R)^2$}\label{Example8_3}
We consider now an example in which fourth, sixth and eight order corrections appear in the action. For the fourth order term we consider a correction of the type $\alpha R^4$ to reduce the number of the parameters involved in the analysis.

For this theory the only non zero auxiliary quantities in Eq.~\rf{XYZT1} are
\begin{align}\label{XYZT1z2}
\begin{split}
& {\bf X}\left(\mathbb{A},\mathbb{R},\mathbb{B}\right)= 6\left(\mathbb{R}+\alpha\frac{6^3   \mathbb{R}^3}{\mathbb{A}^3}+\beta\frac{6^2 \mathbb{B} \mathbb{R}}{\mathbb{A}^2}+\gamma\frac{6\mathbb{B}}{\mathbb{A}^3}\right), \\
&{\bf Y}_1\left(\mathbb{A},\mathbb{R},\mathbb{B} \right)= 1+4 \alpha\frac{ 6^3 \mathbb{R}^3}{\mathbb{A}^3}, 
\\
&{\bf Y}_2\left(\mathbb{A},\mathbb{R} ,\mathbb{B}\right)= \frac{\beta}{\mathbb{A}^2},\\
&{\bf Y}_3\left(\mathbb{A},\mathbb{R} ,\mathbb{B}\right)= 6\left(\beta\frac{ 6 \mathbb{R}}{\mathbb{A}^2}+\gamma\frac{2\mathbb{B}}{\mathbb{A}^3}\right),\\
& {\bf Z}_1\left(\mathbb{A},\mathbb{R},\mathbb{B} \right) = \frac{6^3 \alpha \mathbb{R}^2}{\mathbb{A}^3},\\
& {\bf Z}_5\left(\mathbb{A},\mathbb{R},\mathbb{B} \right) =\frac{2\gamma}{\mathbb{A}^3},\\
&{\bf W}_1\left(\mathbb{A},\mathbb{R} ,\mathbb{B}\right) =\frac{144 \alpha }{\mathbb{A}^2}.
\end{split}
\end{align}
As before, the cosmological equations can be decoupled to give an explicit equation for $\mathbb{S}_3$ and  another for $\mathbb{S}_4$. However, we will not show them here due to their size.

The  dynamical system Eq.~\rf{DynSys8} is now
\begin{align}\label{DynSysRR4BRBR2}
\begin{split}
\DerN{\mathbb{R}}&=-(\mathbb{R}-2)^2+\mathbb{J}+(\mathbb{K}-2) \mathbb{K},\qquad \DerN{\mathbb{A}}=2  \mathbb{A}(\mathbb{K}-\mathbb{R}+2),\\ 
\DerN{\Omega}&=(1-3 w-2 \mathbb{K}+2 \mathbb{R})
   \Omega,\qquad \DerN{\mathbb{K}}=2 \mathbb{K} (\mathbb{K}-\mathbb{R}+1), \\
   \DerN{\mathbb{J}}&=\mathbb{B}+\mathbb{J} (-5 \mathbb{K}+5 \mathbb{R}-3)+22 \mathbb{K}+12,\qquad
   \DerN{\mathbb{S}_1}=\left(2 + \mathbb{K} - \mathbb{R}\right) \mathbb{S}_1 + \mathbb{S}_2,\\\DerN{\mathbb{B}}&=6 \mathbb{K}^4-24 \mathbb{R} \mathbb{K}^3-26
   \mathbb{K}^3+36 \mathbb{R}^2 \mathbb{K}^2+66 \mathbb{R} \mathbb{K}^2-123 \mathbb{K}^2-24 \mathbb{R}^3 \mathbb{K}-54 \mathbb{R}^2
   \mathbb{K}\\
&~~~+226 \mathbb{R} \mathbb{K}-146 \mathbb{K}+6 \mathbb{R}^4+14 \mathbb{R}^3 -4 \mathbb{J}^2-103 \mathbb{R}^2+136 \mathbb{R}+\mathbb{B}  (-3 \mathbb{K}+3 \mathbb{R}+1)\\
&~~~+\mathbb{J} \left(17 \mathbb{K}^2-34 \mathbb{R} \mathbb{K}+36 \mathbb{K}+17 \mathbb{R}^2-34
   \mathbb{R}+37\right)-\mathbb{S}_1-68,\\
\DerN{\mathbb{S}_2}&=\mathbb{S}_2 f_1(\mathbb{R},\mathbb{B},\mathbb{J},\mathbb{S}_1,\mathbb{K}, \A) + \frac{\alpha}{\gamma} f_2(\mathbb{R},\mathbb{B},\mathbb{J},\mathbb{S}_1,\mathbb{K}, \A) \\
&~~~+ \frac{\beta}{\gamma} f_3(\mathbb{R},\mathbb{B},\mathbb{J},\mathbb{S}_1,\mathbb{K}, \A)+  \frac{1}{\gamma}f_3(\mathbb{R},\mathbb{B},\mathbb{J},\mathbb{S}_1,\mathbb{K}, \A)
\,,
\end{split}
\end{align}
where the full equation for $\mathbb{S}_2$ is only shown in its structure due to its length. The full expression can be easily calculated and does not add anything to the understanding of the properties of the dynamical system.

The system in Eq.~\eqref{DynSysRR4BRBR2} presents the usual invariant submanifolds  $\mathbb{A}=0$, $ \Omega=0$, $\mathbb{K}=0$. The fixed points with their stability and the parameter $\mathfrak{s}_4$ that characterise the solution is given in Table \ref{TavolaRR4BRBR2}.

The dynamics of this case is very similar to the one of the previous case, with the difference that the line of fixed points is not present. The only possible attractor is given by one of the points $\mathcal{H}_i$ whereas all the other points are unstable.

\begin{table*}
\begin{center}
\caption{Fixed points of $f= R+ \alpha R^4 +\beta R \Box R + \gamma(\Box R)^2$ and their  associated solutions.
Here A stands for attracotr, R stands for repeller, S stands for saddle, and NHS for non-hyperbolic saddle.  The index ``$i$'' of the points $\mathcal{H}_i$ runs from 1 to 5. The value of $\mathbb{R}^*_i$ are the roots of Eq.~\eqref{8Ord_q_conditon}. }
\label{TavolaRR4BRBR2}
\begin{tabular}{llllll} \hline\hline
Point & Coordinates $\{\mathbb{R},\mathbb{B},\mathbb{J},\mathbb{S}_1,\mathbb{S}_2,\mathbb{K},\Omega, \A\}$ & Solution & Existence&  Stability \\ \hline\\
$\mathcal{A}$ & $\left\{ 1,0,1,1, -1,0, 0 , 0\right\}$  & $\mathfrak{s}_4=-1$ & $\alpha,\beta, \gamma \neq 0$ & NHS \\ \\
$\mathcal{B}$ & $\left\{0,0, 4, 16 , -32, 0, 0, 0\right\}$  & $\mathfrak{s}_4=-128$ & $\alpha,\beta, \gamma \neq 0$ & \makecell[l]{$R$ for $w<1/3$\\$S$ for $w>1/3$}\\ \\
$\mathcal{C}$ & $\left\{2,0, 0,0,0, 0, 0,12\left[\alpha (q-2)\right]^{\frac{1}{q-1}} \right\}$  & $\mathfrak{s}_4=0$ &\makecell[l]{if $q\in\Re$\\ $\alpha(q-2)>0$}& S\\ \\
$\mathcal{I}_1$ & $\left\{24,0, 1,1,-1, 23, 0, 0\right\}$  & $\mathfrak{s}_4=-1$ & $\alpha,\beta, \gamma \neq 0$ & S\\ \\
$\mathcal{H}_i$ & $\left\{\mathbb{R}^*_i , -6\mathbb{R}^*_i(\mathbb{R}^*_i-1)(\mathbb{R}^*_i-2),(\mathbb{R}^*_i-2)^2,(\mathbb{R}^*_i-2)^4,(\mathbb{R}^*_i-2)^5,0,0,0\right\}$  & $\mathfrak{s}_2=\sigma_{i}$  & $\alpha,\beta, \gamma \neq 0$ & \makecell[l]{One A for\\$-0.0044\lesssim \frac{\alpha}{\gamma}\lesssim -0.0060$}\\
  && && \\\hline\\
\multicolumn{5}{c}{\makecell[l]{$\sigma_{i}=\frac{\alpha}{\gamma}\left(-150 \mathbb{R}_{*,i}^4+435 \mathbb{R}_{*,i}^3-252 \mathbb{R}_{*,i}^2\right)+101
   \mathbb{R}_{*,i}^5-610 \mathbb{R}_{*,i}^4+1306 \mathbb{R}_{*,i}^3-1180 \mathbb{R}_{*,i}^2+416 \mathbb{R}_{*,i}-32\neq 0$}}\\ 
 \\\hline\hline\\
 \end{tabular}
   \end{center}
\end{table*}

\section{Analysis of the results}

The structure of the phase space has similarities in all of the particular cases studied. For example, all of those cases feature a fixed point which is a past attractor, that we denoted as point $\mathcal B$. This point is not a global feature of the phase space, as it does not lay in the intersection of all the invariant submanifolds. Also, fixed points $\mathcal A$ and $\mathcal C$ exist in all the cases studied and they are always unstable. 

Concerning the attractors of the theory, we find that for the model of Section~\ref{Example6_1} there exists one global attractor, point ${\mathcal I}_2$. Point ${\mathcal I}_2$ is characterised by $\mathbb{B}\neq 0$, i.e., it represents a state in which the higher-order terms $\Box R$ of the theory are dominant. This is an unexpected result, as it is normally assumed that these terms to be less and less important as the curvature becomes smaller and smaller.  The fact that ${\mathcal I}_2$ is an attractor seems to indicate that instead the cosmology of these theories tends to a state with $\mathbb{B}\neq 0$. Such a state is represented by a solution in which the scale factor converges to a constant value asymptotically. The theory contains a fixed point which can represent a solution with a finite time singularity $\mathcal C$, but this point is always unstable. Since the approach to such solution is very common in theories of fourth order of the form $f(R)$, (but also in $f(\mathcal G)$ theories), this results suggests that six order theories of this type do not incur in these singularities, because the phase space orbits do not converge to fixed points which represent them.

In Section~\ref{Example6_2} we have put at test the robustness of the previous result considering a theory which contains a fourth order term on top of the sixth order one. We have that even in this case in the action the time asymptotic phase space is characterised by a $\mathbb{B}\neq 0$ and therefore to a static universe, while the fixed point $\mathcal C$ is unstable. This is an interesting phenomenon as in Ref.~\cite{Carloni:2015jla} it was shown that points of the type $\mathcal C$ are very often attractors in the phase space for fourth order models. This result suggest that higher order terms might ``cure'' the pathologies induced by the fourth order ones. 

In Section~\ref{6thVS4th} we have  given an explicit analysis of this possibility. In particular, we have shown that the same fixed point $\mathcal C$ appears in the phase space for the theory  $f= R+ \alpha R^{q}$ and  $f= R+ \alpha R^{q} +\gamma R \Box R$.  For the values of the parameter $q$ for which $\mathcal C$  is an attractor the fourth order model, the same point is unstable (saddle). This indicates that the inclusion of sixth order terms is able to prevent the onset of a singularity that would otherwise plague its fourth order  counterpart. In this sense, the sixth-order theory seems to be ``more stable''. The final state of the cosmology, however, depends on the value of $q$. In the specific case $q=3$ this endpoint is represented by one of the points $\mathcal H$. However this is not true for all values of $q$, as the points ${\mathcal H}_i$ do not exist for $q\neq 3$. In this case the final state of the cosmology is probably a point in the asymptotic part of the phase space which we have not explored here.  Clearly we have considered here only a particular example and therefore we cannot prove that this behavior is general. However, the fact that any fourth order model analysed in Ref.~\cite{Carloni:2015jla}  presents one or more points of the type  $\mathcal C$ and that we always expect a change in stability of the corresponding sixth order theory, suggests that we are reporting here a general phenomenon.

The phase space structure is basically the same when one introduces eight order terms. When these terms are added directly to the Hilbert-Einstein Lagrangian the attractor of the new theory is a static cosmology which corresponds to the dominance of the eight order terms. When also fourth order terms are introduced, we observe the same phenomenon observed in the case of sixth order actions: the potentially pathological fixed point that is present in the fourth order gravity phase space becomes unstable for every value of the parameters. We conclude therefore that, like for sixth order terms, also the inclusion eight order therms is able to avoid the onset of singularities. We also considered a model in which fourth, sixth and eight order terms are present in order to estimate their comparative effect. However, in the formalism we have chosen six and eight-order are indistinguishable. A different set of variables might resolve this degeneracy, but its determination and use is left for a future work.

\section{Conclusions}
In this paper we have applied dynamical systems techniques to analyse the structure of the phase space of $f\left(R,\Box R\right)$ gravity. Our choice of dynamical variables allows us to study the cosmology of this entire class of theories by means of a phase space which has at most dimension eight. We have then considered some examples of theories of order six and eight designed specifically to highlight the influence that higher-than-fourth order terms have on the cosmological evolution.  We found that there is complex interplay between terms of different order which make the time asymptotic behavior of these cosmological model non trivial and not easily deducible from their lower order counterpart. Remarkably, we found  that higher order terms can profoundly modify the  behavior of the cosmology, preventing, for example the occurrence of singularities which are known to be induced by the lower order terms \cite{Carloni:2015jla}.  

Connecting our results with the ones available in literature, we can state that
our analysis confirms the result of the absence of a double inflationary phase in theories of order six in full accord with the results of Refs.~\cite{Amendola:1993bg,Gottlober:1989ww}. Indeed, we are able to extend this conclusion also to theories of order eight. This might indicate that no theory of the type $f(R, \Box R)$ is indeed able to generate multiple inflationary phases in spite of their multiple scalar field representation.  Further work on these might be able to confirm this hypothesis.

\appendix
\section{Explicit cosmological and dynamical equations }\label{App}
Cosmological equations for the case \ref{Example6_1}
\begin{equation}
\begin{split}
\mathbb{S}_1&=\mathbb{K} \left[-6\mathbb{B}
+(218-32 \mathbb{J}) \mathbb{R}+38 \mathbb{J}-22 \mathbb{R}^3-52
   \mathbb{R}^2-154\right]+\mathbb{R} [6 \mathbb{B}-34 (\mathbb{J}-4)]\\
   &~~~-6\mathbb{B}-\frac{9 \mathbb{J}^2}{2}+\mathbb{K}^2 \left(16 \mathbb{J}+33
   \mathbb{R}^2+62 \mathbb{R}-121\right)+(16 \mathbb{J}-99) \mathbb{R}^2+41
   \mathbb{J}+\frac{11 \mathbb{K}^4}{2}\\
   &~~~+\mathbb{K}^3 (-22 \mathbb{R}-24)+\frac{11
   \mathbb{R}^4}{2}+14 \mathbb{R}^3-76-\frac{A^2 (\mathbb{K}-\Omega +1)}{12 \gamma },\\
\mathbb{S}_2&=\frac{A^2 (11 \Omega  (\mathbb{R}-\mathbb{K})+\mathbb{K} (11 \mathbb{K}-11
   \mathbb{R}+23)-9 \mathbb{R}+3 w \Omega -13 \Omega +12)}{12 \gamma }\\
   &~~~+\mathbb{K}
   \left[\mathbb{R} (68 \mathbb{B}-864 \mathbb{J}+4086)-26 (4
   \mathbb{B}+97)-\frac{151 \mathbb{J}^2}{2}\right.\\
   &~~~\left.+3 (59 \mathbb{J}-652)
   \mathbb{R}^2+926 \mathbb{J}+\frac{295 \mathbb{R}^4}{2}+46
   \mathbb{R}^3\right]\\
   &~~~+\mathbb{R} \left(100 \mathbb{B}+\frac{151
   \mathbb{J}^2}{2}-864 \mathbb{J}+2264\right)+\mathbb{R}^2 (-34 \mathbb{B}+395
   \mathbb{J}-1815)\\
   &~~~+\mathbb{K}^2 \left[-34
   \mathbb{B}+(2210-177 \mathbb{J}) \mathbb{R}+469 \mathbb{J}-295
   \mathbb{R}^3-138 \mathbb{R}^2-2295\right]\\
   &~~~+5 \mathbb{B} (3 \mathbb{J}-19)+\mathbb{K}^3 \left(59
   \mathbb{J}+295 \mathbb{R}^2+138 \mathbb{R}-830\right)+(576-59 \mathbb{J})
   \mathbb{R}^3\\
   &~~~+5 (89-11 \mathbb{J}) \mathbb{J}+\frac{59 \mathbb{K}^5}{2}+\mathbb{K}^4
   \left(-\frac{295 \mathbb{R}}{2}-46\right)-\frac{59 \mathbb{R}^5}{2}+932,
   \label{a1a1}
\end{split}
\end{equation}

Cosmological equations for the case \ref{Example6_2}
\begin{align}
\begin{split}
\mathbb{S}_1&=3 \alpha  \mathbb{R} \left(6 (\mathbb{J}+(\mathbb{K}-2) \mathbb{K}-4)+(3-9 \mathbb{K})
   \mathbb{R}+4 \mathbb{R}^2\right)+\frac{A^2}{12 \gamma } (\mathbb{K}-\Omega +1)\\
  &~~~\frac{1}{2 \gamma }\left\{\mathbb{K} \left(4 (3 \mathbb{B}-19 \mathbb{J}+77)+(64 \mathbb{J}-436) \mathbb{R}+44
   \mathbb{R}^3+104 \mathbb{R}^2\right)\right.\\
  &~~~+\mathbb{K}^2 \left(-32 \mathbb{J}-66 \mathbb{R}^2-124
   \mathbb{R}+242\right)+\mathbb{K}^3 (44 \mathbb{R}+48)-11
   \mathbb{K}^4\\
  &~~~\left.-4 \mathbb{R} (3 \mathbb{B}-17 \mathbb{J}+68)+(198-32 \mathbb{J}) \mathbb{R}^2-11 \mathbb{R}^4-28 \mathbb{R}^3+12
   \mathbb{B}+\mathbb{J} (9 \mathbb{J}-82)+152\right\}\,,\\
\mathbb{S}_2&=\frac{\alpha}{\gamma} \left\{18
   \mathbb{K}^4+\mathbb{K}^3 (126 \mathbb{R}-72)+\mathbb{K}^2 \left[36
   (\mathbb{J}-2)-387 \mathbb{R}^2-36 \mathbb{R}\right]\right.\\
  &~~~\mathbb{K} \left(18 (7
   \mathbb{J}-52) \mathbb{R}-72 (\mathbb{J}-4)+357 \mathbb{R}^3+99
   \mathbb{R}^2\right)\\
  &~~~\left.+(756-162 \mathbb{J}) \mathbb{R}^2-18 \mathbb{R} (\mathbb{B}-12 \mathbb{J}+48)+39 \mathbb{R}^3-114 \mathbb{R}^4+18 (\mathbb{J}-4)^2\right\}+\\
  &~~~\frac{A^2}{12\gamma} \left\{\frac{11 \mathbb{K}^2}{12}+\mathbb{K} \left(\frac{1}{12} (23-11 \Omega
   )-\frac{11 \mathbb{R}}{12}\right)+\frac{1}{12} \mathbb{R} (11 \Omega -9)+\frac{1}{12}
   (3 w-13) \Omega +1\right\}\\
  &~~~ +\mathbb{K} \left[\mathbb{R} (68 \mathbb{B}-864 \mathbb{J}+4086)-26 (4
   \mathbb{B}+97)-\frac{151 \mathbb{J}^2}{2}+3 (59 \mathbb{J}-652) \mathbb{R}^2+926
   \mathbb{J}+\frac{295 \mathbb{R}^4}{2}+46 \mathbb{R}^3\right]\\
  &~~~+\mathbb{K}^2
   \left(-34 \mathbb{B}+(2210-177 \mathbb{J}) \mathbb{R}+469 \mathbb{J}-295
   \mathbb{R}^3-138 \mathbb{R}^2-2295\right)\\
  &~~~+\mathbb{K}^3 \left(59 \mathbb{J}+295
   \mathbb{R}^2+138 \mathbb{R}-830\right)+\mathbb{K}^4 \left(-\frac{295
   \mathbb{R}}{2}-46\right)+\frac{59 \mathbb{K}^5}{2}\\
  &~~~-\frac{59 \mathbb{R}^5}{2}+(576-59 \mathbb{J}) \mathbb{R}^3+\mathbb{R}^2 (-34 \mathbb{B}+395
   \mathbb{J}-1815)\\
  &~~~ +\mathbb{R} \left(100
   \mathbb{B}+\frac{151 \mathbb{J}^2}{2}-864 \mathbb{J}+2264\right)+5 \mathbb{B} (3 \mathbb{J}-19)+5 (89-11
   \mathbb{J}) \mathbb{J}-932.
   \end{split}
   \label{b1b1}
\end{align}

Cosmological equations for the case \ref{Example8_1}
\begin{align}
\begin{split}
  \mathbb{S}_3&=\mathbb{Q} \left(2 \mathbb{B} (2 \mathbb{J}-\mathbb{K}+6)+646
   \mathbb{J}^2+\mathbb{J} (-78 \mathbb{K}+184 \mathbb{S}+159)-26 \mathbb{K} \mathbb{S}-8
   \mathbb{K}+317 \mathbb{S}+125 \mathbb{S}_1+15 \mathbb{S}_2\right)\\
&~~~+\mathbb{Q}^2
   \left[11 \mathbb{B}-40 \mathbb{J} \mathbb{K}+\mathbb{J} (161 \mathbb{J}+1261)+4
   \mathbb{K}^2-38 \mathbb{K}+507 \mathbb{S}+66 \mathbb{S}_1-108\right]\\
&~~~+\mathbb{Q}^3
   (\mathbb{B}+727 \mathbb{J}-48 \mathbb{K}+97 \mathbb{S}+294)+\mathbb{Q}^4 (43 \mathbb{J}-4 \mathbb{K}+425)+72\mathbb{Q}^5+\mathbb{Q}^6\\
&~~~+38 \mathbb{J}^3+\mathbb{J}^2 (206-18 \mathbb{K})+\mathbb{J}
   \left(4 \mathbb{K}^2-2 \mathbb{K}+194 \mathbb{S}+27\mathbb{S}_1-36\right)\\
&~~~-8 \mathbb{K}^2-10
   \mathbb{K} \mathbb{S}-4 \mathbb{K} \mathbb{S}_1-12 \mathbb{K}+15 \mathbb{S}^2+3 \mathbb{S}+23 \mathbb{S}_1+9 \mathbb{S}_2\\
&~~~+\mathbb{B} (7 \mathbb{J}-2
   \mathbb{K}+\mathbb{S})+\frac{\mathbb{B}^2}{2}+\frac{A^3 \left( \mathbb{R}-\Omega -1-\mathbb{Q}\right)}{12\gamma },\\
\mathbb{S}_4&=\frac{A^3}{12\gamma} \left[-22 \mathbb{Q}^2+11 \mathbb{Q} (2 \mathbb{R}-2
   \Omega -3)+ (2 \mathbb{K}+9 \mathbb{R}-3 (w+4) \Omega -9)\right]\\
&~~~+\mathbb{Q} \left[11 \mathbb{B}^2+2 \mathbb{B} (95 \mathbb{J}-31 \mathbb{K}+11
   \mathbb{S}+54)+340 \mathbb{J}^3+\mathbb{J}^2 (6985-260 \mathbb{K})\right.\\
&~~~+\mathbb{J} \left(72
   \mathbb{K}^2-934 \mathbb{K}+2574 \mathbb{S}+120 \mathbb{S}_1+1251\right)-152
   \mathbb{K}^2-406 \mathbb{K} \mathbb{S}\\
&~~~\left.-44 \mathbb{K} \mathbb{S}_1-232 \mathbb{K}+63
   \mathbb{S}^2+2823 \mathbb{S}+1165 \mathbb{S}_1+145 \mathbb{S}_2\right]\\
&~~~+\mathbb{Q}^2 \left[11 \mathbb{B} (8 \mathbb{J}-4 \mathbb{K}+33)+10543
   \mathbb{J}^2+\mathbb{J} (-1844 \mathbb{K}+2554 \mathbb{S}+13911)\right.\\
&~~~\left.+2 \mathbb{K} (28
   \mathbb{K}-222 \mathbb{S}-291)+8571 \mathbb{S}+2098 \mathbb{S}_1+174
   \mathbb{S}_2-972\right]\\
&~~~+\mathbb{Q}^3 \left[(251 \mathbb{B}-776 \mathbb{J}
   \mathbb{K}+\mathbb{J} (2774 \mathbb{J}+27195)+84 \mathbb{K}^2-1348 \mathbb{K}\right.\\
&~~~\left.+8871
   \mathbb{S}+1026 \mathbb{S}_1+1350\right]+\mathbb{Q}^4 (22 \mathbb{B}+13372
   \mathbb{J}-1024 \mathbb{K}+1711 \mathbb{S}+9483)\\
&~~~+\mathbb{Q}^5 (826 \mathbb{J}-84 \mathbb{K}+8588)+\mathbb{J} \left(2
   \mathbb{K} (28 \mathbb{K}+30 \mathbb{S}-5)+1078 \mathbb{S}-69 \mathbb{S}_1-42
   \mathbb{S}_2-324\right)\\
&~~~+1410 \mathbb{Q}^6+21 \mathbb{Q}^7-4 \mathbb{K}^2 (\mathbb{S}+28)+2\mathbb{K} \left(59
   \mathbb{S}+18 \mathbb{S}_1-2 \mathbb{S}_2+66\right)+52 \mathbb{S}^2+56 \mathbb{S}_1\\
&~~~+63\mathbb{S}+204 \mathbb{S}_1+58 \mathbb{S}_2+\frac{9
   \mathbb{B}^2}{2}+9 \mathbb{B} (7 \mathbb{J}-2
   \mathbb{K}+\mathbb{S})-276 \mathbb{J}^3-\mathbb{J}^2 (142 \mathbb{K}+294
   \mathbb{S}-1779).
   \end{split}
   \label{c1c1}
\end{align}

Dynamical equation for $\mathbb{S}_2$  for the case \ref{Example8_1}
\begin{align}
\begin{split}
\DerN{\mathbb{S}_2}&=\mathbb{S}_2 (16 \mathbb{K}-16
   \mathbb{R}+23)-\frac{29 \mathbb{B}^2}{2}-\mathbb{B}
   \left[16 \mathbb{J} (4 \mathbb{K}-4 \mathbb{R}+9)+71
   \mathbb{K}-53 \mathbb{R}-261\right.\\
&~~~ \left.+(\mathbb{K}-\mathbb{R}) \left(67
   \mathbb{K}^2+\mathbb{K} (259-134 \mathbb{R})+\mathbb{R} (67 \mathbb{R}-225)\right)\right]-38 \mathbb{J}^3\\
&~~~+\mathbb{J}^2 \left(335
   \mathbb{K}^2+\mathbb{K} (780-670 \mathbb{R})+\mathbb{R} (335
   \mathbb{R}-762)+601\right)\\
&~~~-\mathbb{J} \left(-409 \mathbb{K}^4+4 \mathbb{K}^3 (409
   \mathbb{R}-103)+\mathbb{K}^2 (6 (238-409 \mathbb{R}) \mathbb{R}+2576)\right.\\
&~~~\left.+4 \mathbb{K}
   (409 \mathbb{R}^3-405\mathbb{R}^2-1154\mathbb{R}+1135)+
   604\mathbb{R}^3-409 \mathbb{R}^4+2044\mathbb{R}^2)-4216\mathbb{R}\right.\\
&~~~\left.+27
   \mathbb{S}_1+1911\right\}+81 \mathbb{K}^6-2 \mathbb{K}^5 (243
   \mathbb{R}+178)+\mathbb{K}^4 [\mathbb{R} (1215 \mathbb{R}+1624)-2926]\\
&~~~-4 \mathbb{K}^3
   (405 \mathbb{R}^3+734\mathbb{R}^2-2792\mathbb{R}+889)+\mathbb{K}^2 \left(1215 \mathbb{R}^4+2624\mathbb{R}^3-15960\mathbb{R}^2+10688\mathbb{R}\right.\\
&~~~\left.-66
   \mathbb{S}_1+3817\right)+\mathbb{K} \left[3 (44 \mathbb{R}-45) \mathbb{S}_1-2
   (2243
   \mathbb{R}^5+578\mathbb{R}^4-5060\mathbb{R}^3+5354\mathbb{R}^2+3169\mathbb{R}+8566)\right]\\
&~~~+81 \mathbb{R}^6+200 \mathbb{R}^5-2402 \mathbb{R}^4+3576 \mathbb{R}^3-66
   \mathbb{R}^2 \mathbb{S}_1+2537 \mathbb{R}^2-7712
   \mathbb{R}\\
&~~~+(139 \mathbb{R}-37) \mathbb{S}_1+3516+\frac{A^3}{12 \gamma } (\Omega-\mathbb{K} -1)\,.
\end{split}
\end{align}

Dynamical equation for $\mathbb{S}_2$  for the case \ref{Example8_2}
\begin{align}
\begin{split}
\DerN{\mathbb{S}_2}&=\mathbb{S}_2 (16
   \mathbb{K}-16 \mathbb{R}+23)+\frac{\mathbb{A}^3}{12 \gamma } (-\mathbb{K}+\Omega -1)\\
 &~~~-\frac{2^{q-3} 3^{q-2} A^{4-q} \mathbb{R}^{q-2}}{\gamma} \left[(q-1) q (\mathbb{J}+(\mathbb{K}-2)
   \mathbb{K}-4)+q \mathbb{R} (\mathbb{K} (3-2 q)+1)+(q-1)^2 \mathbb{R}^2\right]\\
&~~~-\frac{29 \mathbb{B}^2}{2}-\mathbb{B}
   \left[16 \mathbb{J} (4 \mathbb{K}-4 \mathbb{R}+9)+71
   \mathbb{K}-53 \mathbb{R}-261\right.\\
&~~~ \left.+(\mathbb{K}-\mathbb{R}) \left(67
   \mathbb{K}^2+\mathbb{K} (259-134 \mathbb{R})+\mathbb{R} (67 \mathbb{R}-225)\right)\right]-38 \mathbb{J}^3\\
&~~~+\mathbb{J}^2 \left(335
   \mathbb{K}^2+\mathbb{K} (780-670 \mathbb{R})+\mathbb{R} (335
   \mathbb{R}-762)+601\right)\\
&~~~-\mathbb{J} \left(-409 \mathbb{K}^4+4 \mathbb{K}^3 (409
   \mathbb{R}-103)+\mathbb{K}^2 (6 (238-409 \mathbb{R}) \mathbb{R}+2576)\right.\\
&~~~\left.+4 \mathbb{K}
   (409 \mathbb{R}^3-405\mathbb{R}^2-1154\mathbb{R}+1135)+
   604\mathbb{R}^3-409 \mathbb{R}^4+2044\mathbb{R}^2)-4216\mathbb{R}\right.\\
&~~~\left.+27
   \mathbb{S}_1+1911\right\}+81 \mathbb{K}^6-2 \mathbb{K}^5 (243
   \mathbb{R}+178)+\mathbb{K}^4 [\mathbb{R} (1215 \mathbb{R}+1624)-2926]\\
&~~~-4 \mathbb{K}^3
   (405 \mathbb{R}^3+734\mathbb{R}^2-2792\mathbb{R}+889)+\mathbb{K}^2 \left(1215 \mathbb{R}^4+2624\mathbb{R}^3-15960\mathbb{R}^2+10688\mathbb{R}\right.\\
&~~~\left.-66
   \mathbb{S}_1+3817\right)+\mathbb{K} \left[3 (44 \mathbb{R}-45) \mathbb{S}_1-2
   (2243
   \mathbb{R}^5+578\mathbb{R}^4-5060\mathbb{R}^3+5354\mathbb{R}^2+3169\mathbb{R}+8566)\right]\\
&~~~+81 \mathbb{R}^6+200 \mathbb{R}^5-2402 \mathbb{R}^4+3576 \mathbb{R}^3-66
   \mathbb{R}^2 \mathbb{S}_1+2537 \mathbb{R}^2-7712
   \mathbb{R}\\
&~~~+(139 \mathbb{R}-37) \mathbb{S}_1+3516+\frac{A^3}{12 \gamma } (\Omega-\mathbb{K} -1)\,.
\end{split}
\end{align}


\end{document}